\documentclass[a4paper]{cas-dc}

\usepackage[numbers]{natbib}

\usepackage{times}  
\usepackage{helvet}  
\usepackage{courier}  
\usepackage{graphicx} 
\usepackage{natbib}  
\usepackage{caption}

\usepackage{booktabs}
\usepackage{amsmath}
\usepackage{multirow}
\usepackage{subfigure}
\usepackage{tabularx}
\usepackage{color}

\def\tsc#1{\csdef{#1}{\textsc{\lowercase{#1}}\xspace}}
\tsc{WGM}
\tsc{QE}

\begin{document}
	\let\WriteBookmarks\relax
	\def\floatpagepagefraction{1}
	\def\textpagefraction{.001}
	\let\printorcid\relax 
	
	\shorttitle{Heterogeneous Subgraph Network with Prompt Learning for Interpretable Depression Detection on Social Media}    
	
	\shortauthors{Chen Chen et al.}
	
	\title[mode = title]{Heterogeneous Subgraph Network with Prompt Learning for Interpretable Depression Detection on Social Media}

	\author{Chen Chen}
	\ead{3120004779@mail2.gdut.edu.cn}

        \author{Mingwei Li}
	\ead{3121004145@mail2.gdut.edu.cn} 
	
	\author{Fenghuan Li}
	\cormark[1]
	\ead{fhli20180910@gdut.edu.cn} 
	
	\author{Haopeng Chen}
	\ead{3120004686@mail2.gdut.edu.cn} 
	
	\author{Yuankun Lin}
	\ead{3120004708@mail2.gdut.edu.cn}

	
	\address{School of Computer Science and Technology, Guangdong University of Technology, Guangzhou, China}
	
	\cortext[1]{Corresponding author} 

	
	\begin{abstract}
			Massive social media data can reflect people's authentic thoughts, emotions, communication, etc., and therefore can be analyzed for early detection of mental health problems such as depression. Existing works about early depression detection on social media lacked interpretability and neglected the heterogeneity of social media data. Furthermore, they overlooked the global interaction among users. 
			To address these issues, we develop a novel method that leverages a Heterogeneous Subgraph Network with Prompt Learning(HSNPL) and contrastive learning mechanisms. Specifically, prompt learning is employed to map users' implicit psychological symbols with excellent interpretability while deep semantic and diverse behavioral features are incorporated by a heterogeneous information network. Then, the heterogeneous graph network with a dual attention mechanism is constructed to model the relationships among heterogeneous social information at the feature level. 
			Furthermore, the heterogeneous subgraph network integrating subgraph attention and self-supervised contrastive learning is developed to explore complicated interactions among users and groups at the user level. Extensive experimental results demonstrate that our proposed method significantly outperforms state-of-the-art methods for depression detection on social media.
	\end{abstract}
	
	
	
	\begin{highlights}
		\item An interpretable depression detection method on social media is developed.
		\item The heterogeneity of social data and the interactivity of social users are learned.
		\item Prompt learning can map users’ implicit interpretable psychological symbols.
		\item Heterogeneous attention network can explore feature-level interactions.
		\item Subgraph contrastive learning can discover user-level interactions.
	\end{highlights}
	
	
	\maketitle
	
	\section{Introduction}
	According to the essential facts provided by the World Health Organization Media Centre, an estimated 300 million people across all age groups suffer from depression globally \cite{reddy2012depression}. Depression is the leading cause of disabilities worldwide and a significant contributor to the global burden of diseases. Depression can persist or relapse, severely affecting one's work, study, or daily life abilities \cite{thapar2022depression}. Traditional depression diagnosis and treatment require face-to-face communication with doctors, which hinders identifying potential patients. Furthermore, many people who suspect that they may endure depression are reluctant or fearful to seek professional help at hospitals or clinics due to various factors such as economic, resource, social, and privacy \cite{fried2022revisiting}. Therefore, although there are available treatment programs and drugs, few people receive the treatment, and the existing programs are also difficult to monitor and track people's mental state for massive samples \cite{giuntini2020review}. The widespread social networks such as Twitter have produced massive social media data, capturing people's authentic thoughts, emotions, communication, etc., which offers valuable opportunities to monitor public health issues, particularly depression \cite{malhotra2022deep}. However, effective depression detection models based on complex social networks entail the following challenges: (1) Interpretability: Depression assessment often entails high-risk and life-critical issues, however, existing works on early depression detection are usually end-to-end, so it is essential to ensure sufficient interpretability \cite{zogan2022explainable, han2022hierarchical}; (2) Social networks contain heterogeneous data such as posts and user behaviors, which have different correlations with each other \cite{hamad2021depressionnet}; (3) Interactivity: There are various forms and degrees of interactions among users and groups in social networks.
	
	Existing end-to-end deep learning approaches predominantly prioritize the improvement of classification performance \cite{linardatos2020explainable}. However, depression detection decisions \cite{ali2023enlightening}, often entail high risks and have life-or-death consequences, underscoring the criticality of establishing trust. Therefore, it becomes imperative to understand the factors driving these decisions \cite{ribeiro2016should, du2019techniques}. The tweets posted by users on social platforms often convey evident sentiment signals. For instance, the tweet "I had a gym class today, and I'm so happy" contains "happy", therefore representing outwardly explicit psychological symbols that are relatively easily recognized. However, users may not always employ explicit psychological terminology to express their emotions, insights, and experiences in their tweets. For example, there is no explicit psychological terminology in the tweet "It has been rainy and very wet for several days", which signifies an implicit psychological symbol. It is challenging to identify these implicit symbols based solely on tweets. To address this challenge, the utilization of psychological depression scales proves to be practical and highly interpretable for assessing depression \cite{snaith1993depression}. These scales are selected based on diagnostic criteria for depression and can be leveraged to learn or map the implicit psychological symbols of individuals in a given moment. In this paper, we present a novel prompt learning approach that leverages the understanding capabilities of the large-scale pre-trained language model, employing depression scales to effectively map the implicit depressive symptoms of users at the psychological level. Additionally, we utilize automatic text summarization techniques to compress tweets into concise conclusive descriptions \cite{hamad2021depressionnet}, as well as mine deep semantic features in tweet texts and behavioral features of users that are helpful for depression detection \cite{nor2022sentiment}. We incorporate these diverse features to provide excellent interpretability at both psychological and linguistic levels.
	
	Nowadays, graph neural networks are effective methods to analyze complex relationships between objects in many domains, such as sentiment analysis on social media\cite{jin2021heterogeneous}. Considering the diverse aspects and complicated interactions in social media data, we construct a heterogeneous information network\cite{linmei2019heterogeneous} to integrate the extracted features. Subsequently, we employ a dual attention mechanism, consisting of type-level attention and node-level attention, to learn interactions among different types of nodes at the feature level. Type-level attention can learn the importance of different adjacent node types, and node-level attention can learn the importance of different adjacent nodes, thus enabling more effective aggregation of heterogeneous features, while enriching the semantics and reducing the impact of noise information. After that, considering the social nature of humans, we formulate users and depression detection as subgraphs and subgraph classification problem, respectively. We develop a subgraph attention mechanism to learn complicated interactions among users while developing a self-supervised contrastive learning mechanism that utilizes user subgraphs as positive samples and generates negative samples by perturbing their features. This contrastive learning mechanism enables us to learn the interactions among users and groups at the user level, thereby further enhancing the distinctiveness of users. Consequently, we present a novel approach that leverages a Heterogeneous Subgraph Network with Prompt Learning(HSNPL) and contrastive learning mechanisms. The contributions of this work are described as follows.
	
	\begin{itemize}
		\item We present to model prompt learning to map users' implicit psychological symbols utilizing depression scales and mine a variety of features that may reflect users' depressive state to provide excellent interpretability at the psychological level.
		\item Considering the heterogeneity of social media information, we construct all features as a heterogeneous information network and construct a heterogeneous attention network to explore complicated interactions among different types of information at the feature level.
		\item We develop a subgraph attention mechanism and a self-supervised contrastive learning mechanism to discover the complicated interactions between users and groups at the user level, resulting in more distinctive user representations.
	\end{itemize}

	\section{Related Works}
	\subsection{Depression Detection on Social Media}
	Social media is usually text-based, so most previous works on depression detection were based on text-only information\cite{raimo2022role}. Chiong et al. \cite{chiong2021textual} studied several text preprocessing and text-based feature techniques to propose a method for textual depression detection on social media. Adarsh et al. \cite{adarsh2023fair} normalized the class imbalance, and then proposed a method with a noise label correction technique for textual depression detection on social media. David et al. \cite{losada2020evaluating} evaluated and improved depression-related lexical resources based on context and word embeddings for depression-related information in Reddit text. User-generated content includes diverse information, so some works consider social behaviors and multimodal information. Figueredo et al. \cite{figueredo2022early} considered the importance of emotions in emojis, which was combined with context-independent word embeddings and fusion techniques for depression detection. Ortega et al. \cite{ortega2022revealing} evaluated first-person pronouns as features for personal statements in depression detection on social media. Gui et al. \cite{gui2019cooperative} used GRU to extract text and image features and applied a cooperative multi-agent model for depression detection, addressing the diversity of user-generated content regarding topics and sentiments. Nevertheless, these works often ignored the short-text characteristics of social media data, information interaction, and the interpretability of depression detection.
	
	\subsection{Interpretability of Depression Detection}
	Research on the interpretability of depression detection often adopted attention mechanisms or knowledge-aware methods. Han et al. \cite{han2022hierarchical} considered linguistic features and used metaphorical concept mapping techniques to detect implicit manifestations of depression. Zhang et al. \cite{ijcai2022p725} proposed a method guided by the similarity between the post and psychiatric scale, which can capture risk posts related to the dimensions in the clinical depression scale and provide interpretable diagnostic bases. Zhang et al. \cite{zhang2023depression} devised a deep knowledge-aware depression detection framework, which identified clinically relevant entities by incorporating medical knowledge in the model and considered the dynamic pattern of depression to provide interpretability. Zogan et al. \cite{zogan2022explainable} considered Twitter users' behaviors, topics, and emotions, thereby providing interpretability from the model level based on a hierarchical attention network for depression detection. However, these works only provided better interpretability at the linguistic level and didn't thoroughly learn the mapping between social media data and depressive symptoms at the psychological level.

	\subsection{Heterogeneous Graph Neural Networks}
	
	Recently, people have become increasingly interested in graph neural networks which are effective methods to analyze complex relationships between objects. Lu et al. \cite{lu2020gcan} composed the users into a fully connected graph, then used a graph convolutional network with a collaborative attention mechanism to fuse the source and user representations to deal with social media tasks. However, they used homogeneous graphs with only one type of nodes and edges, and the network structure was relatively simple. Wang et al. \cite{wang2019heterogeneous} proposed a heterogeneous graph attention network with semantic-level and node-level attention to learn the importance of meta-paths and node neighbors, thereby obtaining the final node representation by the aggregation. To address the sparsity/ambiguity and label scarcity problems of short text classification, Hu et al. \cite{linmei2019heterogeneous} proposed a heterogeneous graph attention network to learn short text representations and combined hierarchical attention mechanism to achieve better information aggregation. 
	
	These works demonstrated the feasibility of heterogeneous graph neural networks. For depression detection, Milintsevich et al. \cite{milintsevich2023towards} performed depression diagnosis based on knowledge graphs and text representations. Mihov et al. \cite{mihov2022mentalnet} created ego-networks from user-user interactions (including replies, mentions, and quoted tweets), then merged them into heterogeneous graphs and classified heterogeneous graphs to perform depression detection. However, social media data is more comprehensive to analyze the posted tweets to discover the relationships between users. When considering the combination of tweets and other heterogeneous social information, more potential associations should be discovered. However, many works are based on global structure, thereby some potential associations are difficult to capture, and more delicate substructures need to be considered.
	
	\subsection{Substructures and Contrastive Learning}

	The ability to detect and analyze specific substructures within graphs is crucial for addressing various tasks involving graph-structured data, particularly in the field of social network analysis \cite{chen2020can}. By considering these substructures and exploring the interplay between local and global contexts, we can obtain a more detailed representation of user interactions and capture their distinctive features.  Some works adopted mutual information to measure the relationship between local and global features for graph embedding. Velickovic et al. \cite{velickovic2019deep} learned a node encoder that maximized the mutual information between node representations and corresponding high-level summaries of graphs. Peng et al. \cite{peng2020graph} extended the conventional mutual information from vector space to graph domain, where they computed mutual information from both node features and topological structure. Che et al. \cite{che2021self} used a self-supervised approach to learn graph representations, using an online network to predict the target network. These works applied mutual information to learn relationships between local and global features, but they ignored the subgraph representations within the global scope.

	In the domain of graph learning, the effectiveness of contrastive learning is widely recognized \cite{le2020contrastive}. It is considered an effective approach for uncovering correlations at both the local and global levels. However, it is important to note that currently, only a limited number of studies have explored the application of contrastive learning specifically in the context of depression detection on social media. Yang et al. \cite{yang2022mental} proposed a knowledge-aware mental module based on dot-product attention and employed a supervised contrastive learning module to capture class-relevant features from label information. However, their contrastive learning mechanism was supervised and focused on local features, without explicitly considering the relationship between local and global features.

	Therefore, aiming for the heterogeneity, interactivity, and interpretability of depression detection on social media, we attempt to combine the heterogeneous graph with subgraph contrastive learning and prompt learning to map implicit psychological symbols of users. 
	We believe that various heterogeneous features of users in social networks reflect their potential psychological symbols and are more suitable to be modeled as the heterogeneous graph.
	Moreover, due to massive users and their activities online, the posted information is widely spread, which implies social connections and more latent interactions. These are crucial for depression detection on social media.
	
	\section{The Proposed Model}

	\begin{figure*}[h]
		\centering
		\includegraphics[width=1.0\textwidth]{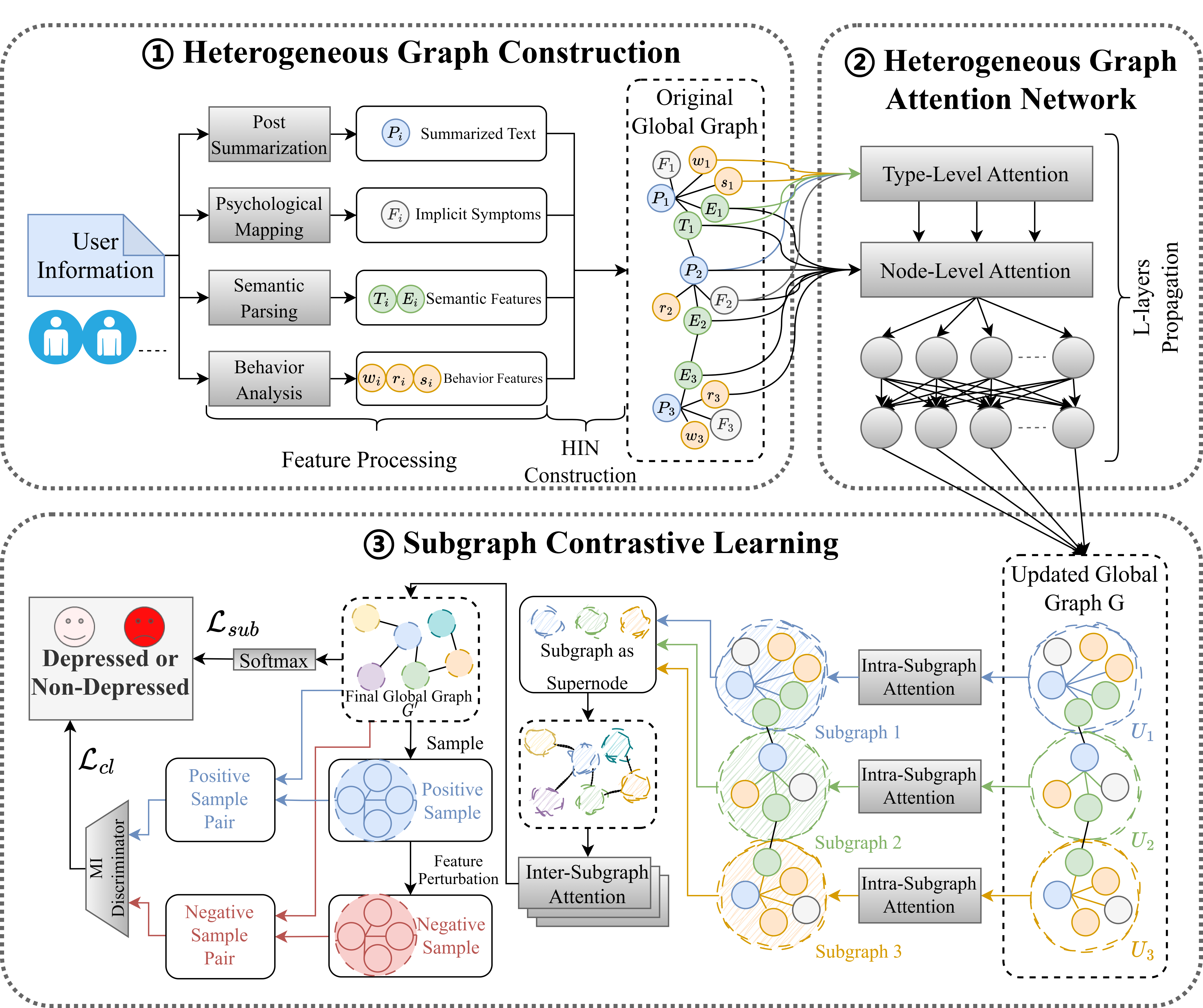}
		\caption{The framework of proposed model.} 
		\label{fig2}
	\end{figure*}
	
	We propose a heterogeneous subgraph network with prompt learning (HSNPL) and contrastive learning mechanisms for depression detection on social media. The framework of our model is shown in Figure \ref{fig2}, which includes three modules. (1) Heterogeneous graph construction: Prompt learning is employed to establish the mapping between social media data and implicit depressive symptoms at the psychological level. Heterogeneous and diverse user information is processed to generate multiple types of nodes, which are then constructed as the original heterogeneous global graph i.e. heterogeneous information network; (2) Heterogeneous graph attention network: The original graph is learned via a heterogeneous graph attention network with a dual attention mechanism to obtain the updated heterogeneous global graph at the feature level; (3) Subgraph contrastive learning: Dual subgraph attention mechanism is employed to learn the interactions between users, then subgraph contrastive learning is developed to learn the interactions between users and groups, therefore, further enhances distinctive subgraph representations at the user level. The final classification loss comprises two components: subgraph classification loss $\mathcal{L}_{sub}$ and contrastive learning loss $\mathcal{L}_{cl}$.
	
	\subsection{Heterogeneous Graph Construction}
	
	\subsubsection{Post Summarization}
	Due to the crucial role of users' posting history in identifying depressive symptoms on social platforms, we analyze their posting histories within one month before their anchor tweets, which are public and large-scale in the dataset \cite{shen2017depression}. However, users post tweets with diverse topics, which may be not relevant or indicative of a depressive state. Therefore, we focus better on the information most related to users' depressive symptoms while reduce the redundancy and noise in the data by means of automatic text summarization techniques. After necessary data preprocessing, we obtain all tweets posted by each user $U_i$ ($i=[1, 2,\ldots, n]$, $n$ is the total number of users) within a period. Given the powerful semantic representation ability of BERT\cite{kenton2019bert}, we employ BERT to learn the embeddings of tweets, followed by clustering the embeddings using the K-means algorithm. The tweets in the cluster centers according to their distances to the centroids are regarded as crucial tweets. For these crucial tweets, we utilize bidirectional and auto-regressive transformers(BART)\cite{lewis2020bart}, which have a bidirectional encoder with a BERT structure and an autoregressive decoder with a GPT structure, to compress potentially redundant information and generate more representative summary text $P_i$. It can automatically select the core features of text content and summarize a large amount of texts into a concise and conclusive description while preserve the critical semantics.
	
	\subsubsection{Psychological Mapping}
	
	\begin{table*}[h]
		\centering
		\caption{The rewritten SDS scale.}
		\begin{tabular*}{\hsize}{@{\extracolsep{\fill}}lcccc}
			\toprule[1pt]
			\midrule
			\multicolumn{1}{c}{Symptoms}  & Rarely & Sometimes & Often & Always \\ \midrule
			1. I {[}mask{]} feel down hearted and blue. & 1 & 2 & 3 & 4 \\
			2. Morning is when I {[}mask{]} feel the best. & 4 & 3 & 2 & 1 \\
			3. I {[}mask{]} have crying spells or feel like it. & 1 & 2 & 3 & 4 \\
			4. I {[}mask{]} have trouble sleeping at night. & 1 & 2 & 3 & 4 \\
			5. I {[}mask{]} eat as much as I used to. & 4 & 3 & 2 & 1 \\
			6. I {[}mask{]} enjoy sex. & 4 & 3 & 2 & 1 \\
			7. I {[}mask{]} notice that I am losing weight. & 1 & 2 & 3 & 4 \\
			8. I {[}mask{]} have trouble with constipation. & 1 & 2 & 3 & 4 \\
			9. My heart {[}mask{]} beats faster than usual. & 1 & 2 & 3 & 4 \\
			10. I {[}mask{]} get tired for no reason. & 1 & 2 & 3 & 4 \\
			11. My mind is {[}mask{]} as clear as it used to be. & 4 & 3 & 2 & 1 \\
			12. I {[}mask{]} find it easy to do the things I used to. & 4 & 3 & 2 & 1 \\
			13. I am {[}mask{]} restless and can't keep still. & 1 & 2 & 3 & 4 \\
			14. I {[}mask{]} feel hopeful about the future. & 4 & 3 & 2 & 1 \\
			15. I am {[}mask{]} more irritable than usual. & 1 & 2 & 3 & 4 \\
			16. I {[}mask{]} find it easy to make decisions. & 4 & 3 & 2 & 1 \\
			17. I {[}mask{]} feel that I am useful and needed. & 4 & 3 & 2 & 1 \\
			18. My life is {[}mask{]} pretty full. & 4 & 3 & 2 & 1 \\
			19. I {[}mask{]} feel that others would be better off if I were dead. & 1 & 2 & 3 & 4 \\
			20. I {[}mask{]} enjoy the things I used to do. & 4 & 3 & 2 & 1 \\  \midrule
			\bottomrule[1pt]
		\end{tabular*}
		\label{tab5}
	\end{table*}
	
	\begin{figure}[h]
		\centering
		\includegraphics[width=1.0\columnwidth]{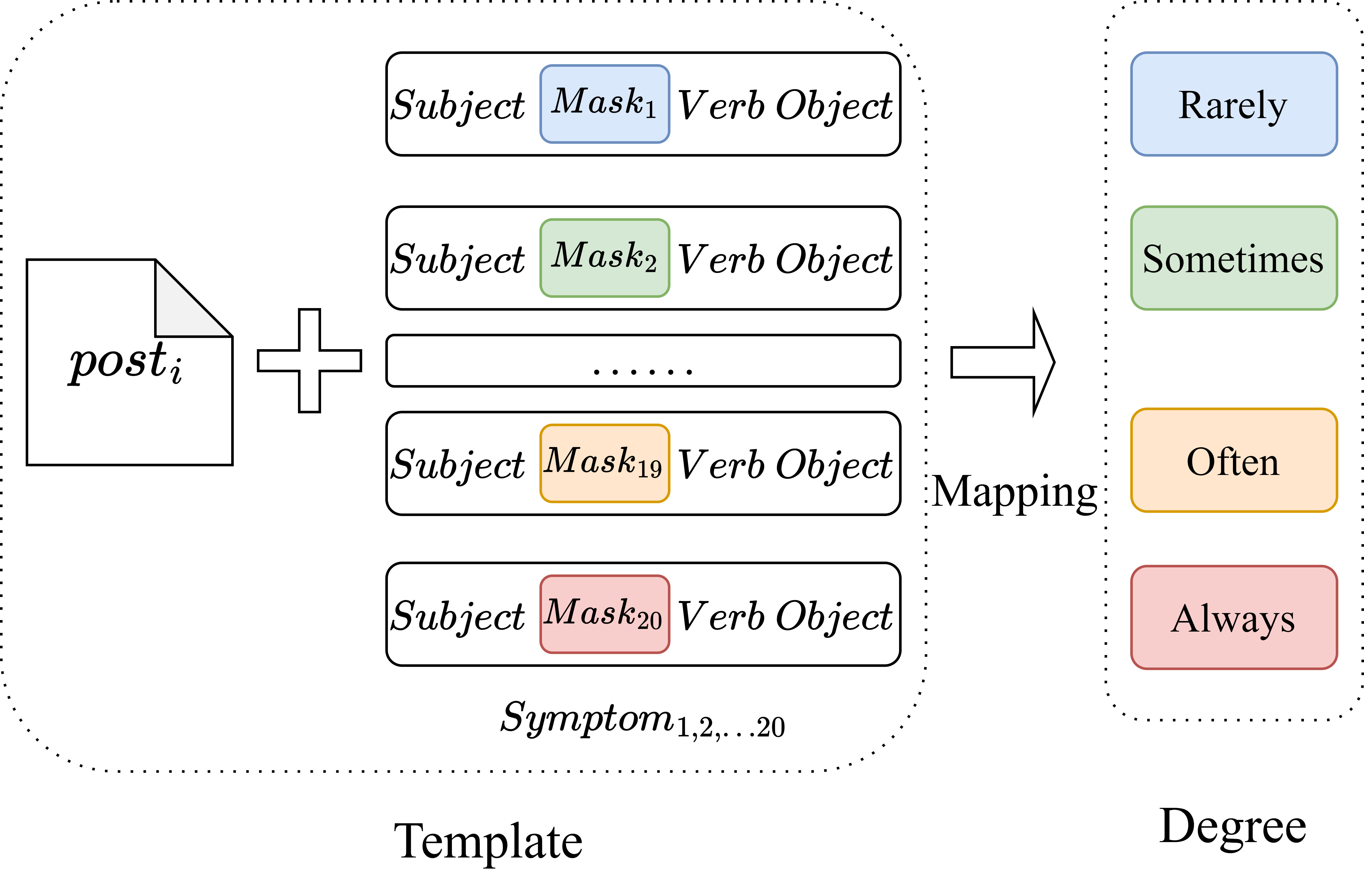}
		\caption{Prompt learning for depressive symptom mapping of posts.}
		\label{fig11}
	\end{figure}

	It is challenging to detect and interpret users' depressive states solely based on their posted tweets. To address this issue, we incorporate the well-known self-rating depression scale(SDS) \cite{zung1965self}, which is a self-report questionnaire with twenty items specifically developed to offer a comprehensive evaluation of various symptoms associated with depression while maintaining brevity, simplicity, and quantifiability.
	We rewrite the symptoms in SDS as "Subject [Mask] Verb Object." as shown in Table \ref{tab5}, where "[Mask]" is an adverb that indicates how often or how much a user experiences a symptom. Next, we use prompt learning which is based on the understanding ability of the large-scale pre-trained language model to map the "[Mask]" of symptoms in SDS for posted tweets. As shown in Figure \ref{fig11}, we first set the template as "post + symptom" where the $post_i$ is defined as all posted tweets by user $U_i$ within a period without summarization. Then, we use a prompt-tuning toolkit \footnote{https://github.com/thunlp/OpenPrompt} to map the "[Mask]" of symptoms to the candidate answers in Table \ref{tab5}, which are \{"Rarely", "Sometimes", "Often", "Always"\}. According to the original SDS, each candidate's answer reflects how severe or frequent a symptom is and is assigned a score from \{1, 2, 3, 4\}. Then, we aggregate the scores of all symptoms according to the candidate answers and obtain a four-dimensional vector $F_i=[F_i^1, F_i^2, F_i^3, F_i^4]$ for each user. Finally, we normalize the vector to get the user's depression scale distribution, which is used as the implicit symptom feature at the psychological level. The formula is as follows:
	
	\begin{equation}\label{eq2}
		F_i^k = \sum_{j=1}^{20}score_{ij}^k
	\end{equation}
	
	Where $score_{ij}^k$ indicates the score corresponding to candidate answer $k$ on $j^{th}$ symptoms that $post_i$ is mapped.
	
	\subsubsection{Semantic Parsing}
	
	Text summarization can filter out much noise and redundant information, but the resulting short texts are also sparse and ambiguous in semantics. Therefore, to obtain more fine-grained text interpretations, we need to discover latent features and their relationships in the texts. To this end, we perform topic analysis on the summarized tweets $P_i$. We use a topic extraction model BERTopic\cite{egger2022topic}, which combines BERT and hierarchical clustering, to identify topics in the summarized tweets $P_i$ and retain the most frequent ones as relevant topics $T_i$. Meanwhile, we perform entity extraction on the summarized tweets $P_i$ utilizing an entity linking tool called TAGME\cite{ferragina2010tagme}, which maps the entities to Wikipedia as well as identifies and disambiguates named entities in each tweet. We obtain the corresponding description texts for each entity and then use BERT to learn their embeddings $E_i$. To capture the semantic relations among the entities, we establish edges between them using cosine similarity measure and only connect the entities whose similarity score exceeds the threshold. Finally, we define the semantic features of the user $U_i$ as $S_i$, composed of two types of nodes: topic and entity, i.e. $S_i= T_i \cup E_i$.
	
	\subsubsection{Behavior Analysis}
	
	\begin{table*}[h]
		\centering
		\caption{Description of user behavioral features.}
		\begin{tabular*}{\hsize}{@{\extracolsep{\fill}}ccc}
			\toprule[1pt]
			\midrule
			Feature                      & Dimension & Description                                                                                                  \\ \midrule
			Time distribution            & 24        & \begin{tabular}[c]{@{}c@{}} the number of tweets posted\\ by a user in each hour.\end{tabular}         \\ \midrule
			Emoticon ratio &
			3 &
			\begin{tabular}[c]{@{}c@{}} the proportion of positive, negative, and\\ neutral emoticons used by a user in his/her tweets \cite{baccianella2010sentiwordnet}.\end{tabular} \\ \midrule
			Sentiment word ratio &
			3 &
			\begin{tabular}[c]{@{}c@{}} the proportion of positive, negative, and\\ neutral sentiment words used by a user in his/her tweets.\end{tabular} \\ \midrule
			Original and retweeted tweets &
			2 &
			\begin{tabular}[c]{@{}c@{}} the number of original\\ and retweeted tweets posted by a user in a period.\end{tabular} \\ \midrule
			Following and follower lists & 2         & \begin{tabular}[c]{@{}c@{}} the number of users\\ that a user follows and is followed by.\end{tabular} \\ \hline
			First-person singular and plural ratio &
			2 &
			\begin{tabular}[c]{@{}c@{}} the proportion of first-person singular\\ and plural pronouns used by a user in his/her tweets.\end{tabular} \\ \midrule
			\bottomrule[1pt]
		\end{tabular*}
		\label{tab4}
	\end{table*}
	
	We consider more features that may reflect users' depressive states on social media, including the time distribution statistics of users' tweets $t_i=\left[t_i^1, t_i^2, \ldots, t_i^{ 24}\right]$, the statistics of original and retweeted tweets $s_i=[s_i^1, s_i^2]$, the statistics of emoticon ratio $e_i=[e_i^1, e_i^2, e_i^ 3]$, the statistics of sentiment word ratio $w_i=[w_i^1, w_i^2, w_i^3]$, the statistics of following and follower lists $f_i=[f_i^1, f_i^2]$, and the statistics of first-person singular and plural ratio $r_i=[r_i^1, r_i^2]$. These multi-dimensional and significant additional features encompass statistics related to the behaviors exhibited in the user's tweets as well as their social interactions. These features represent the actions or behaviors of one user on the social network, whether consciously or unconsciously. By analyzing these behaviors, we can uncover more features associated with users' psychological symbols based on tweets. Detailed descriptions of these behavioral features can be found in Table \ref{tab4}. We normalize these behavioral features respectively and define them as $B_i= t_i \cup e_i \cup w_i \cup s_i \cup f_i \cup r_i$.
	
	These data with multiple types of nodes and edges are typical heterogeneous data, and we model them as a heterogeneous graph structure. To this end, we let the summarized tweets $P_i$ be the central node, and the additional features $O_i = S_i \cup F_i \cup B_i$ be the auxiliary nodes, then $U_i=P_i \cup O_i$ represents all the features of each user. A heterogeneous information network (HIN) is constructed to integrate all users $U=[U_{1}, \ldots, U_{n}]$ and their features. Thus, an original heterogeneous global graph that contains all user features is obtained. 
	
	
	\subsection{Heterogeneous Graph Attention Network}
	The features of users in the original graph vary considerably. It needs to be more comprehensive to model these features, so we must consider the importance of different features. Given a specific node, different types of adjacent nodes may have different impacts on it, and different neighboring nodes of the same type could also have different importance. Therefore, we use a dual attention mechanism to better aggregate features in the heterogeneous information network to capture the different importance at the node and type levels.
	
	Type-level attention can learn the weights of different adjacent node types. Given a specific node $v$, according to the embedding of each adjacent node $v^\prime\in {N_v}$ with type $\tau$, $x_{v^\prime} \in R^q$, where $q$ denotes the length of the feature vector (we perform dimensionality mapping for all features to ensure the same length), we can obtain the embedding of type $\tau$ of its adjacent nodes as $x_\tau=\sum_{v^\prime}{A_{vv^\prime}x_{v^\prime}}$ (where $A$ is the normalized adjacency matrix). Then, we calculate the type-level attention score and use the softmax function to normalize it. The specific formulas are as follows:
	
	\begin{equation}\label{eq6} 
		a_{v,\tau}=\sigma[\mu^T_\tau\cdot(x_v||x_\tau)] 
	\end{equation}
	
	\begin{equation}\label{eq7}
		\alpha_{v,\tau}=\frac{\exp{\left(a_{v,\tau}\right)}}{\sum_{\tau^\prime\in\mathcal{T}}\exp{\left(a_{v,\tau^\prime}\right)}}
	\end{equation}
	
	Where $\mu_\tau$ represents the attention coefficient for type $\tau$, $||$ represents the concatenate, $\mathcal{T}$ is the set of different node types and $\sigma\left(\bullet\right)$ represents the activation function.
	
	Node-level attention can learn the importance of different adjacent nodes and reduce the weights of noisy nodes. Given a specific node $v$ with type $\tau$ and its adjacent node $ v^\prime\in{N_v}$ with type $\tau^\prime$, according to the node embeddings $x_v$ and $x_{v^\prime}$ as well as the type-level attention weight $\alpha_{v,\tau^\prime}$, we can calculate the node-level attention score by multiplying them instead of directly concatenating them, which can make the node-level attention score more reasonably reflect the importance of adjacent nodes with different types and can reduce the dimensionality increase brought by the concatenation operation. Then, we normalize it with the softmax function, where $\widetilde{\mu}$ is the attention coefficient.
	
	\begin{equation}\label{eq8}
		b_{vv\prime}=\sigma[\widetilde{\mu}^T\cdot\alpha_{v,\tau\prime}(x_{v}x_{v\prime})]
	\end{equation}
	
	\begin{equation}\label{eq9}
		\beta_{vv\prime}=\frac{\exp{\left(b_{vv\prime}\right)}}{\sum_{i\in N_v}\exp{\left(b_{vi}\right)}}
	\end{equation}

	Finally, we obtain the propagation rule for the heterogeneous global graph as Formula (\ref{eq10}). After L-layer propagation for the original graph, we obtain updated heterogeneous global graph representation $G$.
	
	\begin{equation}\label{eq10}
		G^{\left(l+1\right)}=\sigma\left(\sum_{\tau\in\mathcal{T}} B_\tau\cdot G_\tau^{\left(l\right)} \cdot W_\tau^{\left(l\right)}\right)
	\end{equation}
	
	Where $B_\tau$ is the attention matrix (the element in row $v$ and column $v^\prime$ is $\beta_{vv^\prime}$), $G_\tau^{\left( 0\right)}=X_\tau$ ($X_\tau$ is the initial feature matrix of all nodes with type $\tau$), $W_\tau^{\left(l\right)}$ represents the transformation matrix of type $\tau$ in the $l^{th}$ layer.
	
	\subsection{Subgraph Contrastive Learning}
	\label{subgraph}
	As users on social media, in addition to constructing a heterogeneous global graph at the feature level, it is crucial to consider the interactions among users.
	For this purpose, we construct each user as one subgraph and consider the finer-grained subgraph structure at the user level. Specifically, we do not change any graph structure and adopt intra-subgraph attention within the subgraph of each user $U_i$ to obtain the impact of each node on this subgraph, thereby achieving the subgraph embedding $g_i$ for user $U_i$ as follows:

		\begin{equation}\label{eq11}
			g_i=\sum_{x_{ij} \in {U_i}} \sigma\left(\alpha_{intra} W_{intra} x_{ij}\right)x_{ij}
		\end{equation}
	
	Where $x_{ij}$ is the embedding of each node belonging to $U_i$, $\alpha_{intra}$ is the attention vector, and $W_{intra}$ is the transformation's weight matrix. 
	
	It is worth noting that there are shared nodes and edges among supernodes, which ensure their interconnectivity and contribute to the formation of user groups. We consider that users within the same group have stronger correlations and closer social activities. To construct subgraph interaction, we treat each subgraph in the global graph as a supernode and use multi-head inter-subgraph attention to learn attention coefficients to distinguish the mutual influence among subgraphs, which reflect the interaction among users. Then we obtain the subgraph embedding $sg_i$ after inter-subgraph attention for user $U_i$ as follows:

		\begin{equation}\label{eq12}
			sg_i=\dfrac{1}{M}\sum_{m=1}^M \sum_{j\in N_i}\beta_{ij}^m W_{inter}^m g_j
		\end{equation}
	
	Where $N_i$ is the set of neighbors of supernode $sn_i$, $\beta_{ij}$ is the attention coefficient of supernode $sn_j$ on supernode $sn_i$, $W_{inter}^m$ is the transformation's weight matrix, and $M$ is the number of attention heads.
	
	Furthermore, to further enhance the discriminative representation of the subgraph as well as the correlation between the subgraph and global graph, which reflect the relationship between users and the groups formed by their shared nodes and edges, we develop a self-supervised contrastive learning mechanism for subgraphs and global graph. In addition to taking existing subgraphs as positive samples, we also need to construct some negative samples(subgraphs). Our specific method is to retain the structure of positive samples but randomly shuffle the features of positive samples to obtain a perturbation as negative samples. This perturbation can be understood as the situation in which both positive and negative sample users engage in various activities on social platforms, such as tweeting, retweeting, and following other users. However, the content of the negative sample users' activities is intentionally perturbed. For instance, their tweets, number of retweets, number of followers, etc. differed from those of positive sample users. Consequently, they exhibit different social interactions with other users or groups on the platforms. Then, the positive sample pair is composed of the positive sample subgraph and the global graph. Similarly, the negative sample pair is composed of the negative sample subgraph and the global graph.
	Based on this self-supervised contrastive learning mechanism, we use mutual information to measure the correlation between the subgraph and global graph, which is optimized by Jensen-Shannon MI estimator \cite{mrn2019learning}, to determine whether the subgraph is from this global graph as follows.
	
	\begin{equation}\label{eq13}
		\mathcal{D}(sg_i,G^\prime)=\sigma\bigl(sg_\mathrm{i}^\mathrm{T}W_{\mathrm{MI}}G^\prime\bigr)
	\end{equation}
	
	Where $W_{\mathrm{MI}}$ is the rating matrix of mutual information, $G^\prime$ is the final global graph representation achieved by pooling on all subgraph embeddings, and $\sigma\left(\bullet\right)$ is the sigmoid function.
	
	Next, we use binary cross-entropy to define the contrastive learning loss, which measures how well the subgraph embeddings match the global embedding. The specific formula is as follows:
	
	\begin{equation}\label{eq14}
		\begin{aligned}
			\mathcal{L}_{cl}=\dfrac{1}{n_{pos}+n_{neg}}\bigl(\sum_{i=1}^{n_{pos}}E_{pos}\bigl[ \log\bigl(\mathcal{D}(sg_{i},G^\prime)\bigr)\bigr] \\ +\sum_{j=1}^{n_{neg}} E_{neg}\left[\log\left(1-\mathcal{D}\left({sg}^{\prime}_j,G^\prime\right)\right )\right])
		\end{aligned}
	\end{equation}
	
	Where $n_{pos}$ denotes the number of positive samples, $n_{neg}$ denotes the number of negative samples, $sg_i$ is the embedding of the positive sample, ${sg}^{\prime}_j$ is the embedding of the negative sample.
	
	\subsection{Training}
	
	Finally, we perform subgraph classification as depression detection, which can determine the user's depression state based on the subgraph containing all features of this user. Furthermore, we consider that the posted texts on social media may contain crucial features that contribute to depression detection. Therefore, we construct the loss of subgraph classification as follows: 
	
	\begin{equation}\label{eq17}
		\begin{aligned}
			\mathcal{L}_{sub}=\sum_{i=1}^n Y_{i}\cdot\{\log[\:softmax(sg_i)]+
			\\
			\log[\:softmax(hp_{i})]\}
		\end{aligned}
	\end{equation}
	
	Where $sg_i$ is the subgraph embedding of one user, $Y_i$ is the corresponding label, $n$ is the total number of users in the dataset, $hp_{i}$ represents the embedding of post node $P_{i}$.
	
	The overall loss function of the model is described as follows: 
	\begin{equation}\label{eq15}
		\mathcal{L}=\alpha\mathcal{L}_{cl}+\beta\mathcal{L}_{sub}+\eta||\Theta||_2
	\end{equation}
	
	Where $\alpha$ and $\beta$ control the contrastive learning and subgraph classification, and $\eta$ is the coefficient for L2 regularization on model parameters $\Theta$. We use gradient descent and an early stopping strategy to perform the optimization.
	
	\section{Experiment Analysis}
	In this section, we conduct extensive experiments and compare our model with some state-of-the-art baselines that use the same dataset to evaluate the performance.
	
	\subsection{Dataset and Experimental Settings}
	
	A large-scale public dataset\cite{shen2017depression} for depression detection on Twitter is used for experiment analysis in this paper, which consists of the following three parts:
	\begin{itemize}
		\item D1: Depressed dataset, where the users who posted anchor tweets containing "(I'm / I was / I am / I've been) diagnosed depression" were labeled as depressed (positive). This part contains 2558 users and 1.2 million tweets.
		
		\item D2: Non-depressed dataset, where the users who never posted tweets containing the word "depression" were labeled as non-depressed (negative). This part contains 5304 users and 2.6 million tweets. We randomly sampled a subset of these users to balance the number of positive and negative users in our dataset. 
		
		\item D3: Depression candidate dataset, where a large-scale unlabeled depression candidate dataset was constructed with 58810 users and 29 million tweets. This part was not used in our work because it was not annotated, and it was unclear how the candidates were selected or filtered. However, we plan to explore this part in our future work as a potential source of data augmentation.
	\end{itemize}
	
	The raw data often contains much noise, which can reduce the usefulness and validity of the data. Therefore, we perform the following data preprocessing to filter out irrelevant contents and extract key information from the raw data.
	\begin{itemize}
		\item We remove the anchor tweets because they could introduce bias or leakage to the depression detection task. We also remove any retweets or duplicate tweets from the same user to avoid data redundancy.
		
		\item We only consider English users and exclude users who have too many followers or too few tweets because they could be bots or spammers. We use a language detection tool to identify the language of each user based on their most recent tweets.
		
		\item We filter out components that appear frequently but are insignificant for depression analysis, such as stopwords, URLs, mentions, and replies. We use a predefined list of stopwords and regular expressions to remove these components from each tweet.
	\end{itemize}
	
	Finally, we obtain a dataset with 5285 users, including 2522 depressed users from D1 and 2763 non-depressed users from D2. The average number of tweets per user is 732, and the average number of words per tweet is 14. Moreover, the parameter settings are shown in Table \ref{tab3}, and 5-fold cross-validation is conducted for experimental analysis. Our experiments are performed on a Tesla P40 GPU with 24 GB of memory, and the computation time is about 150 minutes.
	
	\begin{table}[h]
		\centering
		\caption{Parameter setting.}
		\resizebox{.7\columnwidth}{!}{
			\begin{tabular}{cc}
				\toprule[1pt]
				Parameter                               & Value  \\ \midrule
				\# Topics                                   & 15     \\
				\# Entities                                  & 1076   \\
				\# Subgraph attention heads                 & 6     \\
				\# Heterogeneous graph layers               & 512    \\
				Bert embedding size                         & 768    \\
				Negative sampling rate           & 1.0    \\
				Entity connection rate                      & 0.5   \\
				Learning rate                           & 0.01   \\
				Dropout rate                                & 0.8    \\
				Momentum                                & 0.8    \\
				Batch size                              & 64     \\
				\# Epoches                                   & 1000   \\ 
				\bottomrule[1pt]
			\end{tabular}
		}
		\label{tab3}
	\end{table} 
	
	\subsection{Baselines}
	The diagnostic information in the tweets has a powerful guiding effect on depression detection. Therefore, it is our main task to perform experiments with the data after filtering out the anchored tweets, but we also compare our model with some methods that used the anchored tweets. The baseline methods are as follows:
	\begin{itemize}
		\item HAN(2017)\cite{yazdavar2017semi}: A word-level hierarchical attention neural network framework was used to analyze user tweets for depression detection;
		
		\item BERT(2018)\cite{kenton2019bert}: A bidirectional transformer encoder (Base) was used to obtain tweet-only embeddings and perform depression detection;
		
		\item RoBERTa(2021)\cite{zhuang2021robustly}: Robustly optimized BERT pre-training approach was applied for tweet-oriented depression detection;
		
		MentalBERT(2021)\cite{ji2022mentalbert}: A Pre-trained BERT model specifically designed for the mental healthcare domain, was applied for tweet-oriented depression detection;
		
		\item ChatGLM(2022)\cite{du2022glm}: ChatGLM is a bilingual (English and Chinese) pre-trained language model. We take it as the pre-trained model, then utilize LoRA \cite{hu2022lora} to fine-tune our task;
		
		\item COMMA(2019)\cite{gui2019cooperative}: A multi-agent reinforcement learning method was used to extract features from texts and images, then GRU and VGG-Net were combined for classification;
		
		\item SenseMood(2020)\cite{lin2020sensemood}: Based on CNN and BERT, deep features were extracted from user-posted images and texts, and then visual and textual features were combined to detect the user's depressive state;
		
		\item DepressionNet(2021)\cite{hamad2021depressionnet}: User's behavioral information on social platforms was considered, and automatic summarization was used to condense text information, then a cascaded deep network was used to connect different levels of behavioral features;
		
		\item MCNN(2022)\cite{zogan2022explainable}: Based on MLP and CNN, a hybrid model that used texts and user profile features was built for depression analysis;
		
		\item MDHAN(2022)\cite{zogan2022explainable}: Based on user behavioral features, a hierarchical attention mechanism was developed for encoding at the word level and sentence level;
		
		\item HAN-MCM(2022)\cite{han2022hierarchical}: A model was proposed based on a hierarchical attention mechanism, which took texts and metaphorical concept mapping as the input to achieve interpretability.
	\end{itemize}
	
	\subsection{Performance Comparison}
	
	\begin{table*}[h]
		\centering
		\caption{Performance comparison of depression detection.}
		\begin{tabular*}{\hsize}{@{\extracolsep{\fill}}ccccccc}
			\toprule[1pt]
			\midrule
			Tweets & Feature & Model & P & R & F1 & ACC \\ \midrule
			\multirow{2}{*}[-2.6ex]{\begin{tabular}[c]{@{}c@{}}with\\ Anchor Tweet\end{tabular}} & Text + MCM & HAN-MCM & 0.975 & 0.969 & 0.972 & 0.971 \\ \cmidrule{2-7} 
			& \textbf{\begin{tabular}[c]{@{}c@{}}Text + Depression Scale\\ + Additional Features\end{tabular}} &  \textbf{HSNPL(OURS)} & \textbf{0.989} & \textbf{0.970} & \textbf{0.979} & \textbf{0.978} \\  \midrule
			\multirow{9}{*}[-5ex]{\begin{tabular}[c]{@{}c@{}}without\\ Anchor Tweet\end{tabular}} & \multirow{5}{*}{Text} & HAN & 0.870 & 0.840 & 0.839 & 0.844 \\
			&  & BERT & 0.877 & 0.801 & 0.866 & 0.832 \\
			&  & RoBERTa & 0.902 & 0.837 & 0.868 & 0.851 \\ 
			&  & MentalBERT & 0.908 & 0.895 & 0.906 & 0.878 \\ 
			&  & ChatGLM & 0.886 & 0.823 & 0.854 & 0.859 \\ \cmidrule{2-7} 
			& \multirow{2}{*}{Text + Image} & SenseMood & 0.903 & 0.870 & 0.886 & 0.884 \\
			&  & COMMA & 0.900 & 0.901 & 0.900 & 0.900 \\ \cmidrule{2-7} 
			& \multirow{3}{*}{\begin{tabular}[c]{@{}c@{}}Text \\ + \\ Additional Feature\end{tabular}} & MCNN & 0.874 & 0.870 & 0.870 & 0.871 \\
			&  & MDHAN & 0.902 & 0.892 & 0.893 & 0.895 \\
			&  & DepressionNet & 0.909 & 0.904 & 0.912 & 0.901 \\ \cmidrule{2-7} 
			& Text + MCM & HAN-MCM & - & - & 0.928 & - \\ \cmidrule{2-7} 
			& \textbf{\begin{tabular}[c]{@{}c@{}}Text + Depression Scale\\ + Additional Features\end{tabular}} & \textbf{HSNPL(OURS)} & \textbf{0.938} & \textbf{0.962} & \textbf{0.950} & \textbf{0.949} \\ \midrule
			\bottomrule[1pt]
		\end{tabular*}
		\label{tab1}
	\end{table*}
	
	\begin{table*}[h]
		\centering
		\caption{The role of each module in the model.}
		\begin{tabular*}{\hsize}{@{\extracolsep{\fill}}cccc}
			\toprule[1pt]
			\midrule
			Modules & Processing & F1 & ACC \\ \midrule
			\multirow{2}{*}{Heterogeneous graph} & w/o Dual Attention & 0.897 & 0.894 \\
			& w/o HIN & 0.840 & 0.834 \\ \midrule
			\multirow{3}{*}{Subgraph} & w/o Contrastive Learning & 0.916 & 0.910 \\
			& w/o Subgraph Attention & 0.902 & 0.898 \\
			& \begin{tabular}[c]{@{}c@{}}w/o Contrastive Learning \& Subgraph Attention\end{tabular} & 0.822 & 0.823 \\ \midrule
			\multirow{3}{*}{Prompt Learning} & w/o Prompt Learning & 0.922 & 0.916 \\
			& \begin{tabular}[c]{@{}c@{}}w/o Heterogeneous graph \& Prompt Learning\end{tabular} & 0.809 & 0.811 \\
			& \begin{tabular}[c]{@{}c@{}}w/o Subgraph \& Prompt Learning\end{tabular} & 0.795 & 0.796 \\  \midrule
			\textbf{HSNPL(Total)} & - & \textbf{0.950} & \textbf{0.949} \\ \midrule
			\bottomrule[1pt]
		\end{tabular*}
		\label{tab2}
	\end{table*}
	
	We conduct experiments on the dataset with and without anchor tweets and categorize the methods according to different features. The results in Table \ref{tab1} demonstrate the effectiveness of our method in improving the performance of depression detection by incorporating text, user behavior, and depression scale as the heterogeneous graph as well as applying subgraph contrastive learning. As shown in Table \ref{tab1}, our method outperforms the existing state-of-the-art methods on all evaluation metrics. 
	
	Tweet-oriented depression detection methods that solely rely on text features, such as HAN and BERT, have been found to have unsatisfactory performance. However, approaches like RoBERTa, which is fine-tuned based on this foundation, and MentalBERT, which is specifically pre-trained for mental healthcare, have shown improved performance. Despite the favorable performance of MentalBERT in the field of mental healthcare, it is one pre-trained language model and downstream tasks on social media should be developed, moreover, we acknowledge that it shares a common limitation with other tweet-oriented approaches, which often fail to consider the implicit symbols at psychological level and may not adequately address the issue of interpretability.
	Notably, ChatGLM does not perform satisfactorily after fine-tuning our task, because it requires more excellent expertise and contextual understanding for the downstream tasks. SenseMood and COMMA use both text and image features, improving the model’s ability to predict depression status, but they ignore the interaction between users. As the dataset only contains image links rather than image content, image retrieval is considered dataset augmentation. Therefore, we do not use image features. DepressionNet and MDHAN consider more dimensions of features and provide interpretability from the model structure. However, they do not integrate these multidimensional features well and fail to discover the imperceptible interactions among them efficiently. HAN-MCM integrates linguistic knowledge in both cases (with and without anchor tweets), achieving superior performance. However, it also does not consider the interactions among social users, which are also crucial for depression detection.

	\subsection{Ablation Study}
	We conduct ablation experiments on modules and features in the model and analyze them from these two aspects.
	
	\subsubsection{Module}
	
	We analyze the role of each module in our model, as shown in Table \ref{tab2}. First, we study the effect of the heterogeneous graph module. The results show that the model's performance would significantly decrease if we remove the heterogeneous graph module and only use the subgraph part with prompt learning to classify each subgraph. This suggests that modeling social data as a heterogeneous graph is crucial for depression detection. Furthermore, if we remove the dual attention mechanism, which captures the interactions among nodes, the model's performance would also decrease significantly, indicating that the dual attention mechanism is reasonable. The correlation among nodes should also be considered, which can capture the features and correlations of social media data. 
	
	Next, we discover the vital role of the subgraph module in our method. If we remove both contrastive learning and subgraph attention mechanisms, the model's performance will drastically deteriorate, showing that the interaction between users and the group is indispensable. Finally, we estimate the contribution of prompt learning to our method. The results indicate that prompt learning can enhance our final performance by around 3$\%$, demonstrating that prompt learning is effective in revealing implicit psychological symbols of users via the depression scale. In addition, if we do not use prompt learning and only retain the heterogeneous graph(w/o Subgraph \& Prompt Learning) or the subgraph part(w/o Heterogeneous graph \& Prompt Learning), the model's performance would decline by almost 15.5$\%$ or 14.1$\%$, respectively, which suggests that subgraph classification provides superior performance to post-node classification and prompt learning can better provide discriminative depressive features at the psychological level.
	
	\subsubsection{Feature}
	
	\begin{figure}[h]
		\centering
		\includegraphics[width=1.1\columnwidth]{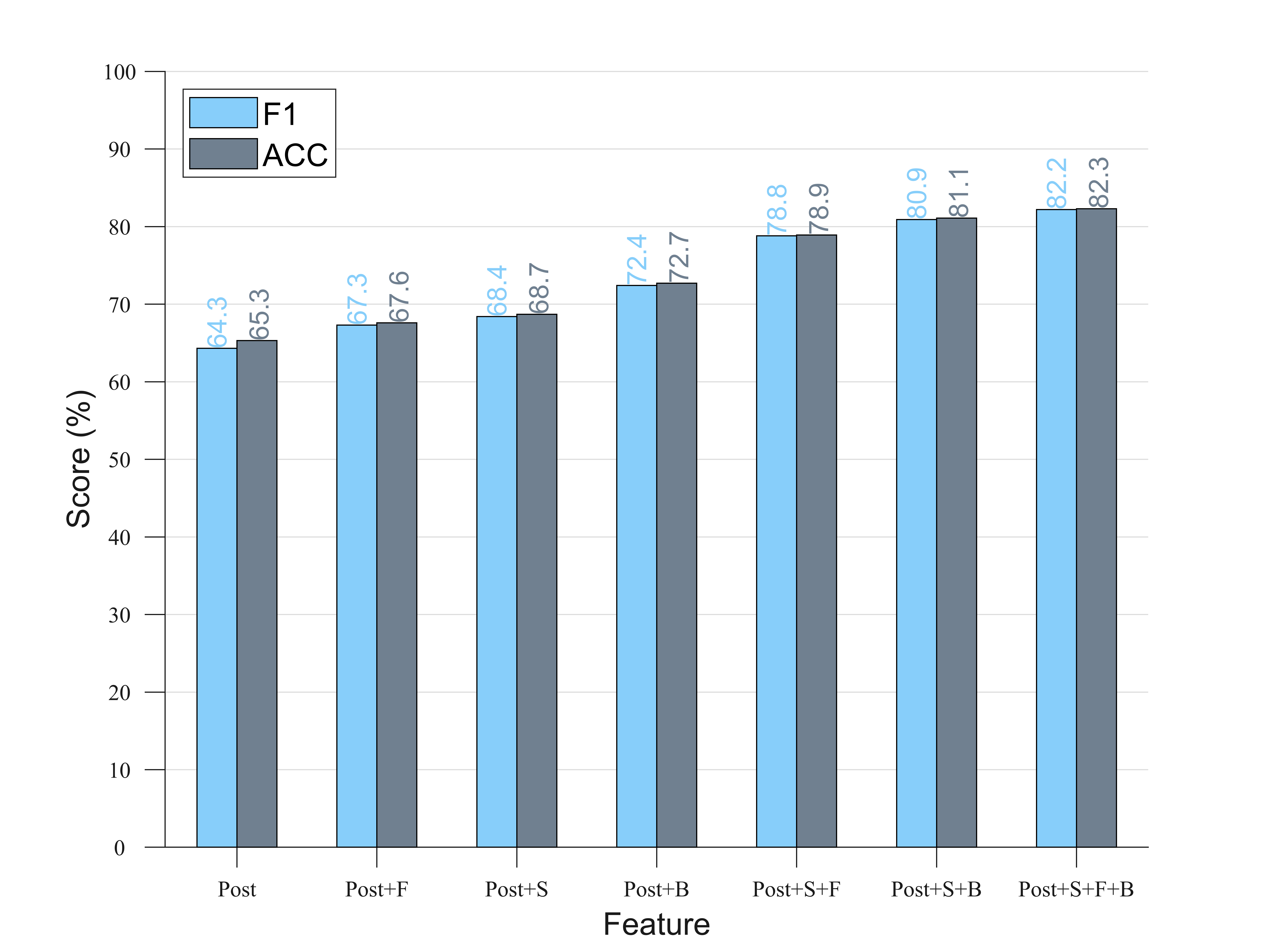}
		\caption{The importance of each additional information.}
		\label{fig6}
	\end{figure}
	
	In this section, we analyze the role of each type of feature in our model, as shown in Figure \ref{fig6}, where $Post$, $F$, $S$, and $B$ denote the textual post after summarization, implicit psychological symptoms, the set of semantic features, and the set of behavioral features, respectively. The results show that in addition to the posts published by users, implicit psychological symptoms mapped by prompt learning, semantic features extracted from the posts, and user behavioral features improve the performance by almost 3$\%$, 4$\%$, and 8$\%$, respectively. In addition, the behavioral and implicit psychological features further improved the performance after aggregating textual posts with semantic features. Furthermore, if we only use textual posts as the features, the performance would decrease by 18$\%$, which indicates that more features beyond the textual posts need to be mined for depression detection on social media. 
	
	We note that behavioral features include six features, and semantic features include topics and entities. Although implicit psychological features are only composed of depression scale scores, they also play a significant role, which highlights the importance of implicit psychological features for depression detection and enhances interpretability. We randomly select five questions from Table \ref{tab4} and report the number of users corresponding to each answer by prompt learning in Table \ref{tab6}, in (number1, number2). "number1" represents the number of depressed users, and "number2" represents the number of non-depressed users. The results reveal a marked difference between the number of depressed and non-depressed users corresponding to the same answer to the same question, which confirms that prompt learning is effective in capturing implicit psychological symbols of users via the depression scale.
	
	\subsection{Interpretability by Prompt Learning}
	
	\begin{table*}[t]
		\centering
		\caption{The reasonability of prompt learning by means of SDS to capture implicit psychological symbols of users.}
		\begin{tabular*}{\hsize}{@{\extracolsep{\fill}}lcccc}
			\toprule[1pt]
			\midrule
			\multicolumn{1}{c}{Symptom \textbackslash \ Degree} & Rarely & Sometimes & Often & Always \\ \midrule
			I {[}mask{]} feel down hearted and blue. & 2239, 9 & 248, 329 & 22, 2386 & 13, 39 \\
			I {[}mask{]} have trouble sleeping at night. & 2304, 448 & 208, 504 & 5, 1757 & 5, 54 \\
			I {[}mask{]} enjoy sex. & 34, 46 & 262, 2345 & 2210, 351 & 16, 21 \\
			I {[}mask{]} get tired for no reason. & 273, 16 & 2184, 147 & 61, 2188 & 4, 412 \\
			I am {[}mask{]} restless and can't keep still. & 1514, 28 & 923, 1186 & 58, 1451 & 27, 98 \\ \midrule
			\bottomrule[1pt]
		\end{tabular*}
		\label{tab6}
	\end{table*}
	
	\begin{table*}[h]
		\centering
		\small  
		\caption{Prompt learning for depressive symptom mapping.}
		\begin{tabularx}{0.9\linewidth}{llXX}
			\toprule[1pt]
			\midrule
			User & Post & Symptom                                          & Degree    \\  \midrule
			\multirow{4}{*}{Depressed} &
			\multirow{4}{*}{\begin{tabular}[c]{@{}l@{}}1. Bereaved families can plant snowdrops in\\ remembrance of loved ones at annual walk.\\ 2. These essential oils can help you when dealing\\ with depression through a bereavement.\\ 3. 10 things I wish someone would have told me \\ about \#grief.\#bereavement \#counselling \#support.\\ 4. \#JohnTravolta posts touching tribute about his \\ \#bereavement after son's death.\\ 5. Coping with the loss of a loved one. \end{tabular}} &
			1. I {[}mask{]} feel down hearted and blue. &
			often \\ \cmidrule{3-4} 
			&      & 2. I {[}mask{]} feel hopeful about the future. & rarely    \\ \cmidrule{3-4} 
			&      & 3. I {[}mask{]} have crying spells or feel like it.      &  always   \\ \midrule
			\multirow{4}{*}{\begin{tabular}[c]{@{}c@{}}Non\\ Depressed\end{tabular}} &
			\multirow{4}{*}{\begin{tabular}[c]{@{}l@{}}1. Today I feel that I want to learn and fun and \\ happy. I LOVE TODAY.\\ 2. Tomorrow I hope it rains because I want to play \\ in the class room with Rita.\\ 3. Today I do it science i'm do it little be it help.\\ 4. Today I had P.E boy v.s girl on doge ball castle \\ it was fun!\\ 5. Today I will have an airsoft gun Fusheng.\end{tabular}} &
			1. I {[}mask{]} feel down hearted and blue. &
			sometimes \\ \cmidrule{3-4} 
			&      & 2. I {[}mask{]} feel hopeful about the future.      & often     \\ \cmidrule{3-4} 
			&      & 3. My life is {[}mask{]} pretty full.               & always    \\  \midrule
			\bottomrule[1pt]
		\end{tabularx}
		\label{tab7}
	\end{table*}
	
	\begin{figure*}[h]
		\centering
		\subfigure[The Number of Topics $K$]{
			\label{fig10.1}
			\includegraphics[width=1.0\columnwidth]{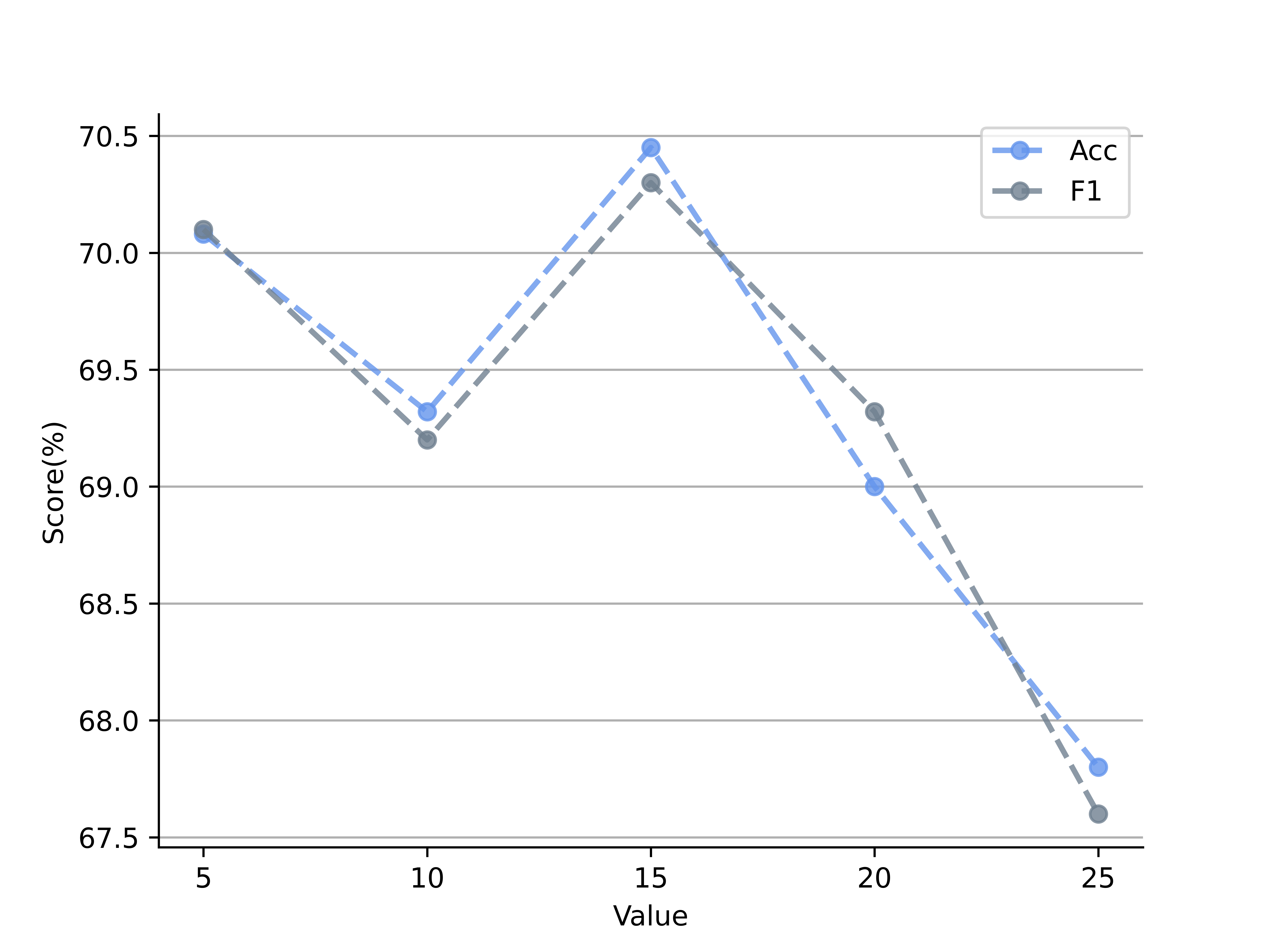}}
		\subfigure[The Number of Entities $E$]{
			\label{fig10.2}
			\includegraphics[width=1.0\columnwidth]{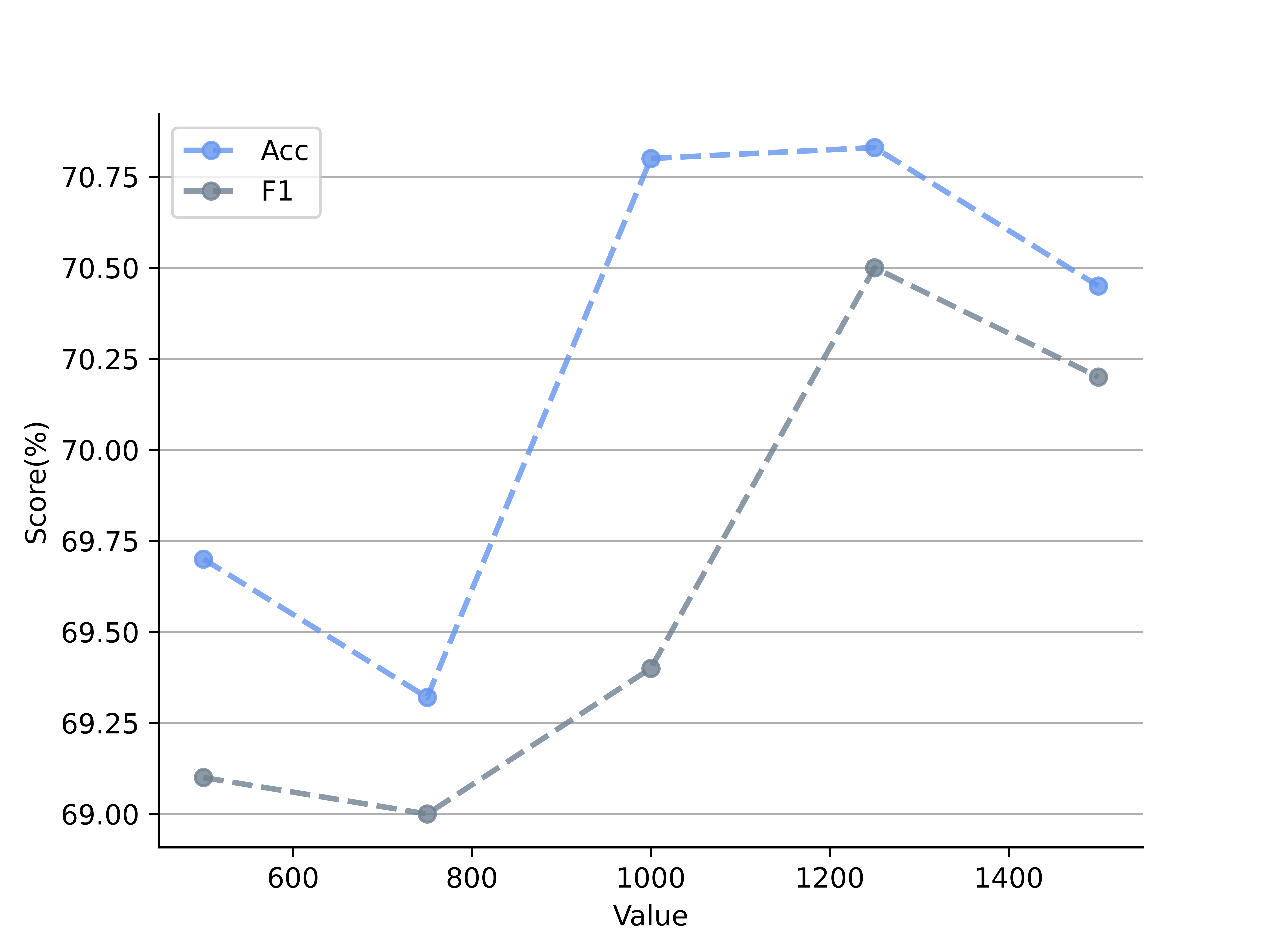}}
		\caption{Parameter Analysis.}
		\label{fig10}
	\end{figure*}
	
	We randomly select one depressed user and one non-depressed user from the dataset to illustrate our method and show the results of prompt learning to map their tweets to depression symptoms, as shown in Table \ref{tab7}. The results show that for users with different depressive states, their tweets are mapped to different answers to the same question by means of prompt learning. For example, for the symptom "I {[}mask{]} feel down hearted and blue.", we use the prompt for depressed and non-depressed users and map their tweets to the degree "often" and "sometimes" respectively, showing that our method can capture subtle differences in psychological symbols between users with different depressive states. For the symptom "I {[}mask{]} feel hopeful about the future.", their tweets are mapped to the degree "rarely" and "often" respectively, which shows that our method can reflect contrasting attitudes toward the future between users with different depressive states. 
	
	We also reveal the ability of prompt learning to map tweets to relevant answers to different questions. For example, for the symptom "I {[}mask{]} have crying spells or feel like it.", the tweets of depressed users are mapped to the degree "always". In contrast, for the symptom "My life is {[}mask{]} pretty full.", the tweets of non-depressed users are mapped to the degree "always". This result shows that our method can identify the distinctive depression symptoms of users with different depressive states, indicating that the prompt learning approach utilizing a depression scale in our method is feasible. It can analyze implicit psychological symbols from users' tweets on social media and obtain more interpretable psychological mapping.
	
	\subsection{Parameter Analysis}
	
	We conduct parameter analysis to investigate the effect of several vital parameters on the model performance, measured by F1-score and ACC. We analyze two parameters, the number of topics $K$ and entities $E$, as shown in Figure \ref{fig10.1} and Figure \ref{fig10.2}. We find that the number of topics $K$ and the number of entities $E$ have an optimal range for the model performance, which is affected by the trade-off between the complexity and the expressiveness of the graph structure. When $K$=15 and $E$=1250, the model can achieve the best performance.
	
	\subsection{Visualization}
	We visualize the information of our method from several aspects: attention mechanism, heterogeneous graph, and contrastive learning.

	\subsubsection{Attention Mechanism}
	\label{Attention}
	
	\begin{figure*}[h]
		\centering
		\subfigure[Dual attention of nodes]{
			\label{fig3}
			\includegraphics[width=1.0\columnwidth]{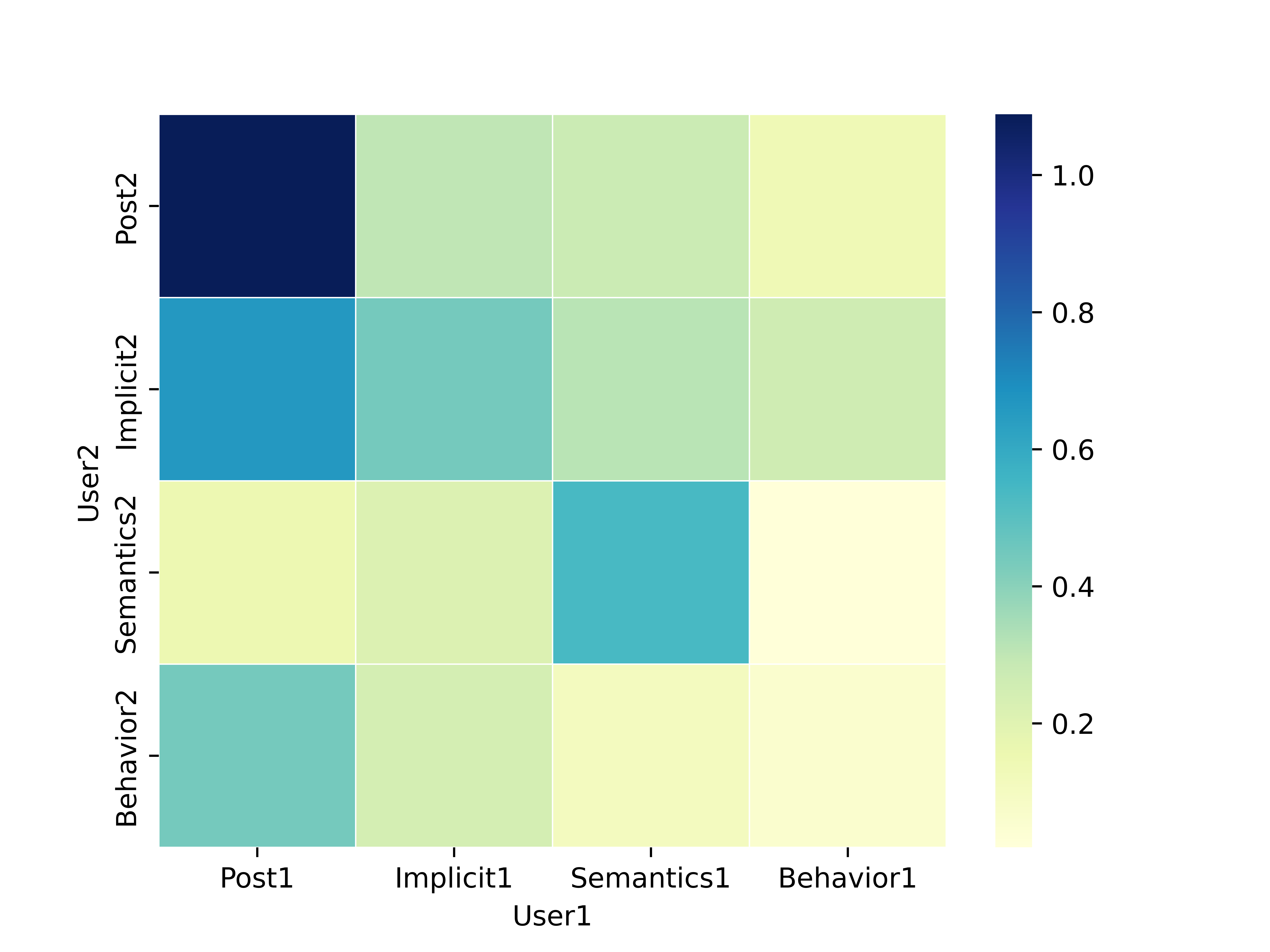}}
		\subfigure[Subgraph attention of users]{
			\label{fig4}
			\includegraphics[width=1.0\columnwidth]{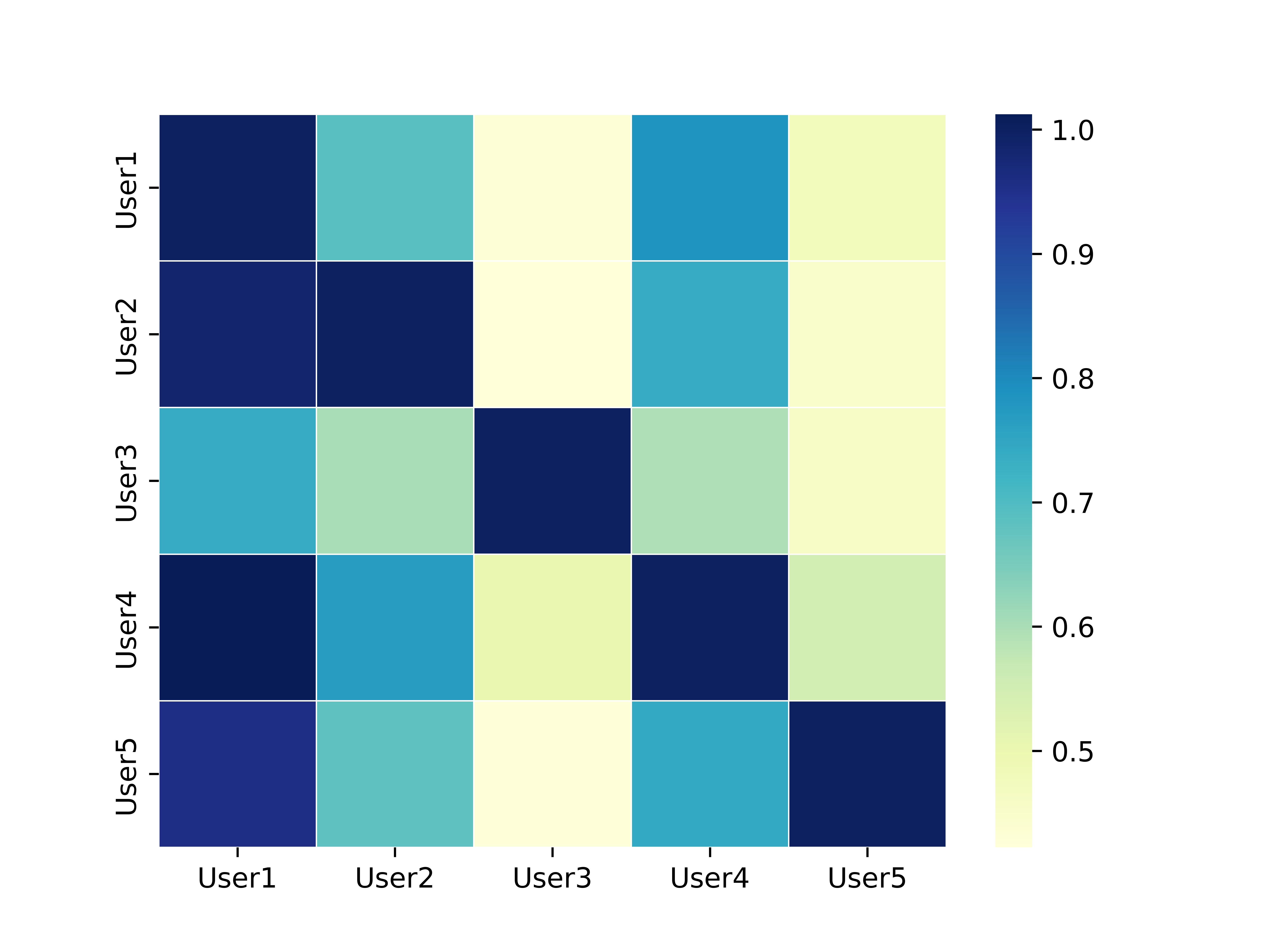}}
		\caption{Visualization of different attention mechanisms.}
	\end{figure*}
	
	\begin{figure*}[h]
		\centering
		\subfigure[Original distribution]{
			\label{fig12}
			\includegraphics[width=0.65\columnwidth]{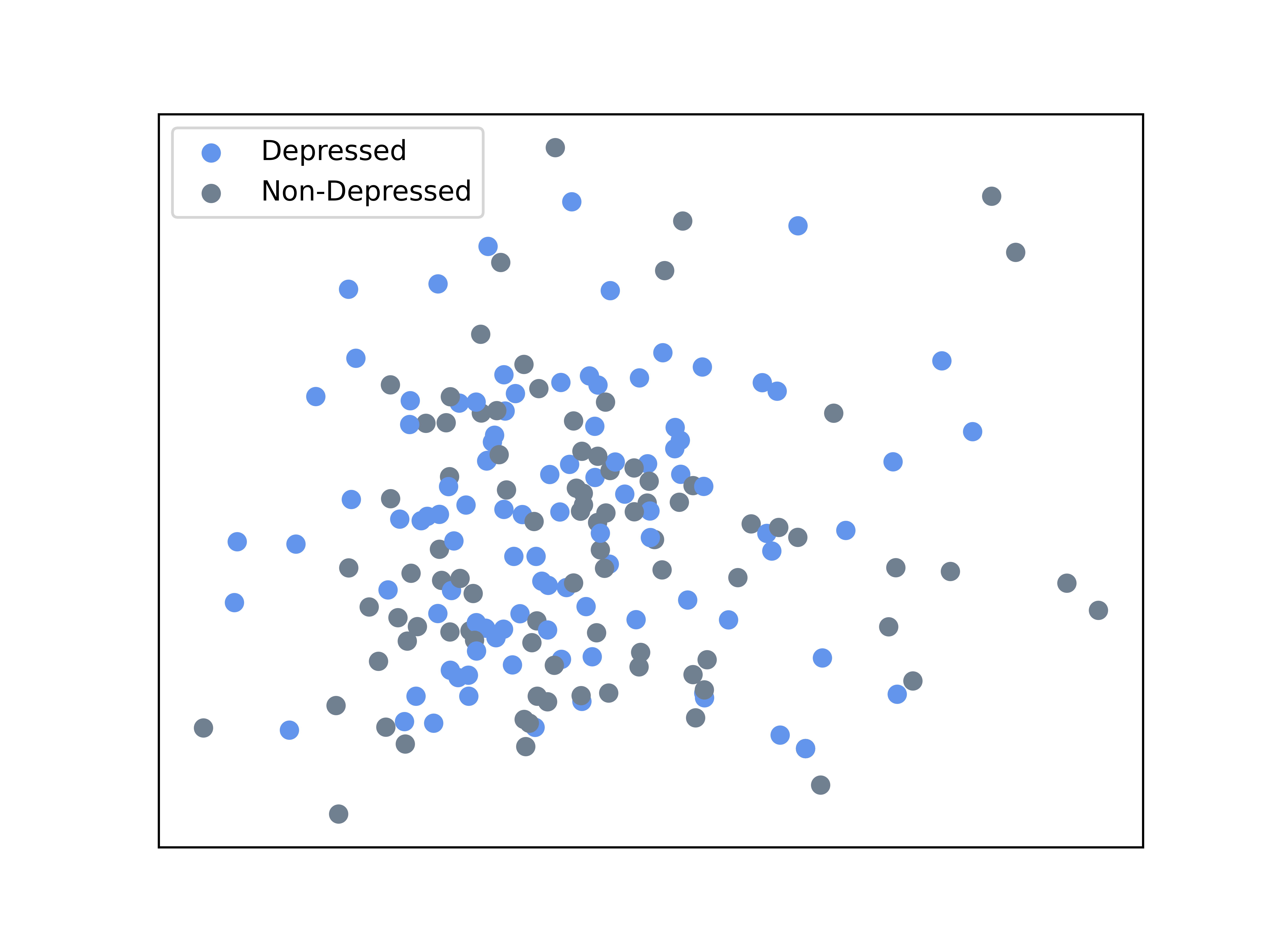}}
		\subfigure[With heterogeneous graph]{
			\label{fig5.1}
			\includegraphics[width=0.65\columnwidth]{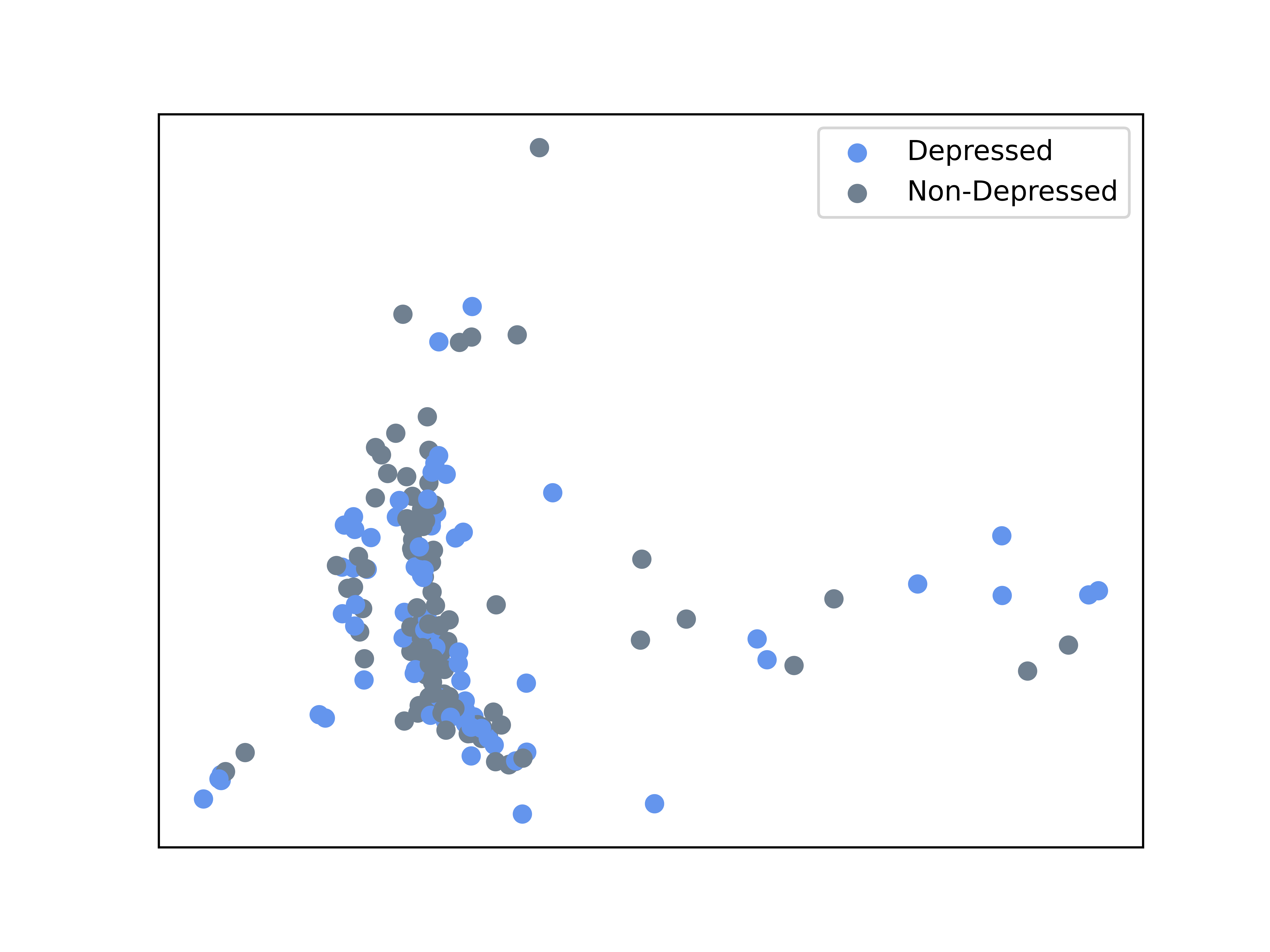}}
		\subfigure[With subgraph contrastive learning]{
			\label{fig5.2}
			\includegraphics[width=0.65\columnwidth]{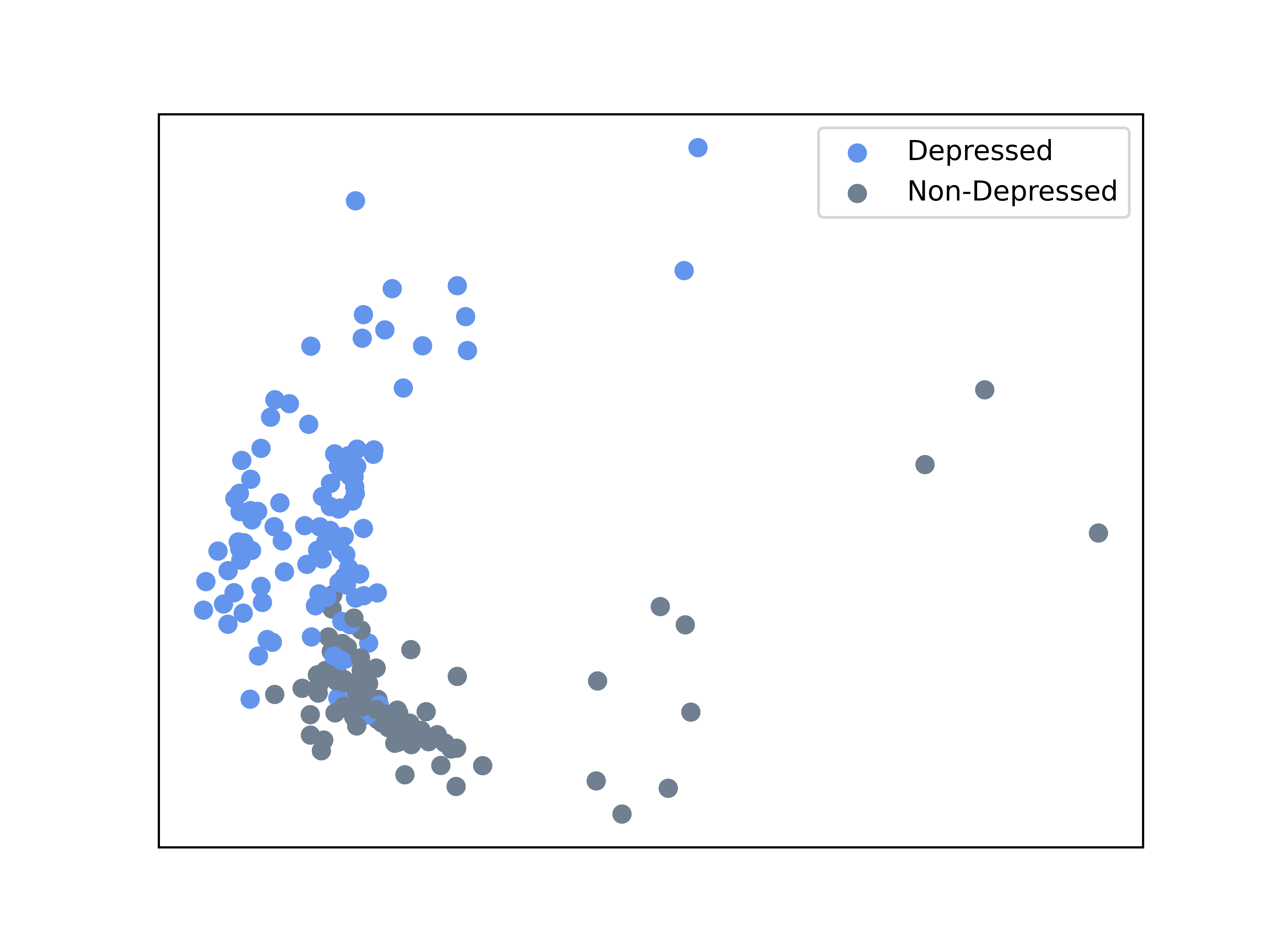}}
		\caption{User feature distribution in our method}
	\end{figure*}
	
	\begin{figure*}[h]
		\centering
		\subfigure[Attention weights obtained by Bert]{
			\includegraphics[width=1.0\columnwidth]{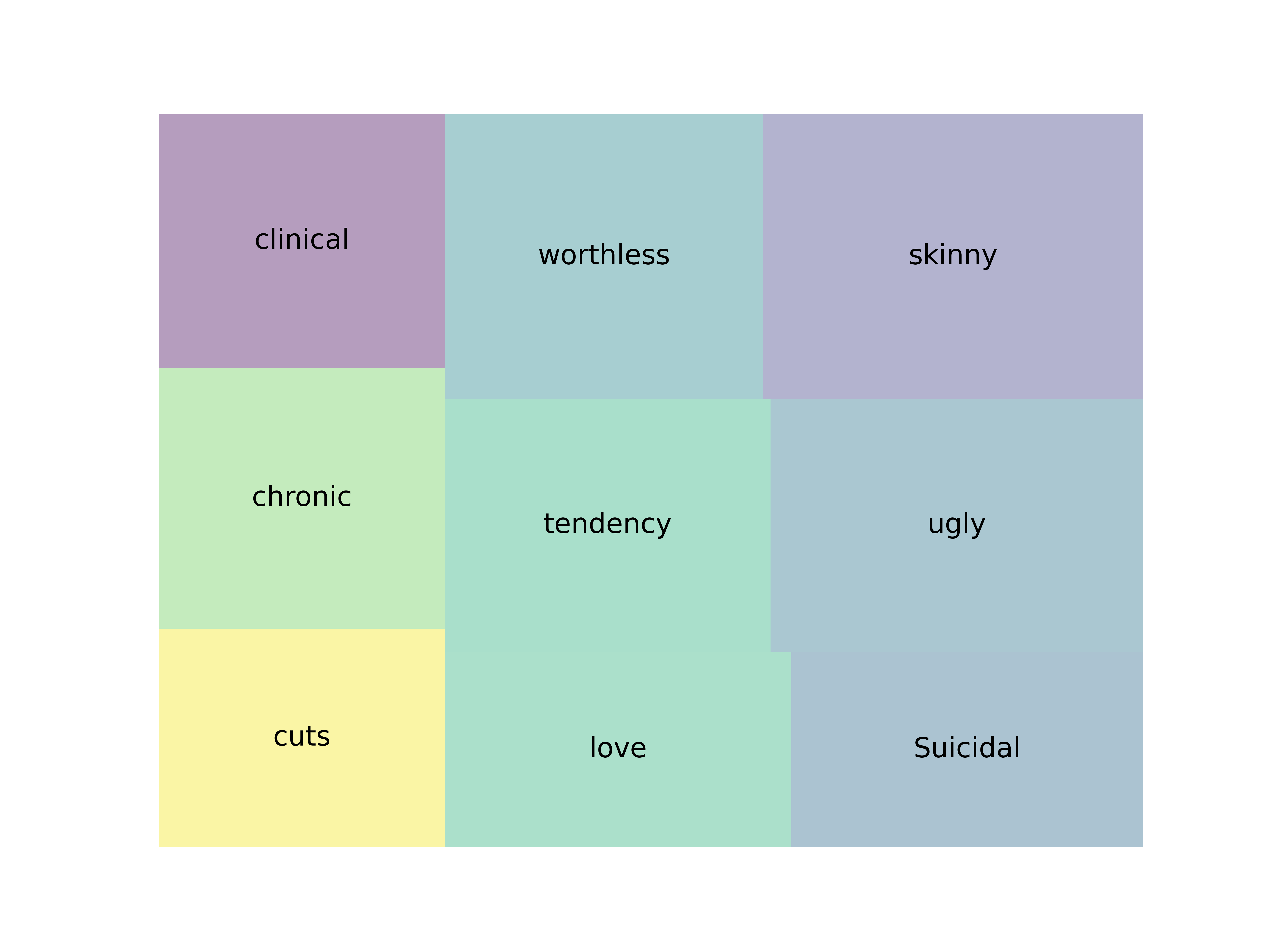}}
		\subfigure[Attention weights obtained by our model]{
			\includegraphics[width=1.0\columnwidth]{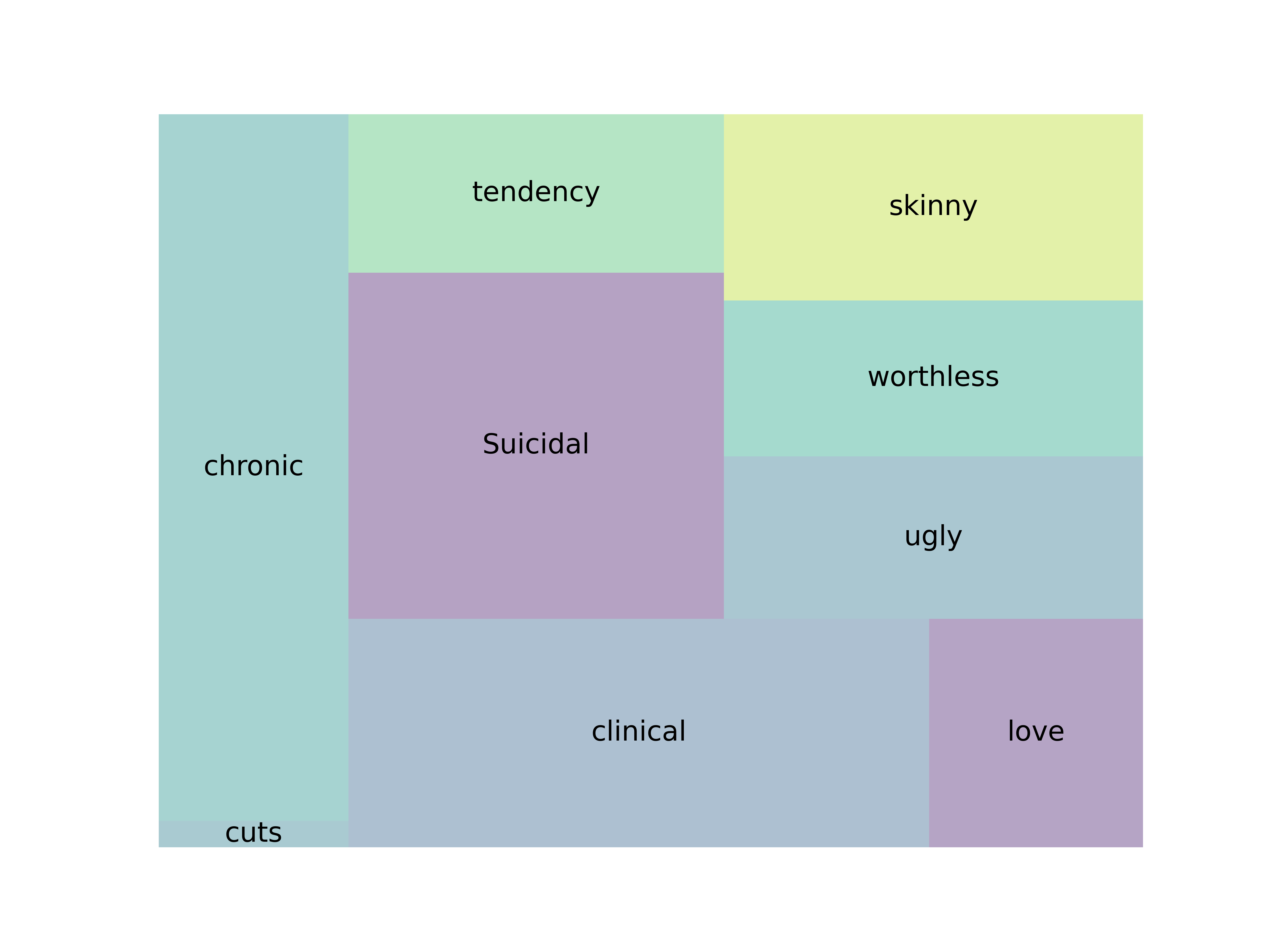}}
		\caption{Attention weights of words obtained by different models.}
		\label{fig13}
	\end{figure*}
	
	We visualize the impact of dual node and subgraph attention mechanisms on the model at both feature and user levels. Firstly, as shown in Table \ref{tab2}, the model's performance drops by about 5$\%$ when either attention mechanism is removed, indicating that merely modeling social information as a heterogeneous graph is insufficient, and the interactions between features and users are essential in predicting depression risk. Secondly, to explain the importance of dual attention mechanism at the feature level in detail, we select two users from the heterogeneous graph and visualize their node attention matrix, as shown in Figure \ref{fig3}, where the horizontal axis represents the nodes of $User_1$, the vertical axis represents the nodes of $User_2$, and the darker the color, the stronger the interaction influence of the horizontal axis node on the vertical axis node. The figure shows that the average attention weight of the $Post_1$ node on the nodes of $User_2$ is the highest, followed by implicit symptom and semantics nodes, and the lowest is the behavioral node, indicating that posts are the core nodes. There are unique edges among semantic nodes and texts, whose relevance is strong, so semantic nodes are assigned more weight. It is worth noting that the implicit symptom node is also assigned much weight, reflecting our method's excellent performance of prompt learning. Finally, since behavioral features are unique to each user and only are constructed relations with the current user, they are assigned fewer weights. Therefore, we believe that dual attention is essential for capturing the interaction among different features.
	
	Then, we select five users who are learned by means of the subgraph attention mechanism in Section \ref{subgraph} and visualize the attention weights among users, as shown in Figure \ref{fig4}, where each block represents the interaction influence of one user in the horizontal axis on one user in the vertical axis. The color represents the weight of the influence. From the figure, we can see that since $User_1$ posted 28$\%$ more tweets than the average number of tweets posted by the users in the dataset, he has the most decisive influence on the other four users in the figure. In addition, $User_1$, $User_2$, and $User_4$ are labeled as depressed users in the dataset. They are assigned more weight than $User_3$ and $User_5$ who are labeled as non-depressed users. The reason may be that depressed users tend to post more negative content, which is more likely to spread on social media, such as anxiety spread, so depressed users are more influential to other users. Moreover, the interaction between two users is different, so the attention weight matrix should be asymmetrical.
	
	\subsubsection{Heterogeneous Graph and Subgraph Contrastive Learning}
	
	To visualize the initial feature distribution of users, we use PCA to reduce feature representations of 200 randomly selected users to two-dimensional space, as shown in Figure \ref{fig12}. Then, we visualize the feature representations of users after heterogeneous graph propagation, as shown in Figure \ref{fig5.1}. From the figure, we can see that the initial user feature representations are chaotic. Nevertheless, the heterogeneous graph can enhance the correlations among different types of features and the interactions among users by modeling these features as a heterogeneous graph structure, thereby improving the effectiveness of depression detection. 
	
	However, after heterogeneous graph propagation, we can also see many overlapping points and outliers in the user distribution, which needs to be improved for high-precision decision-making. We illustrate the contribution of subgraph contrastive learning on user feature representations, as shown in Figure \ref{fig5.2}. We can see that by means of this module, user representations are more clearly distinguished, and the model achieves a better discrimination ability for depressed and non-depressed users, indicating that this module can obtain more distinctive user representations by learning the interactions between users and the group. Therefore, it plays a crucial role in our method.
	
	\subsection{Case Study}
	
	\begin{table*}[h]
		\centering
		\caption{Case Study.}
		\begin{tabular*}{\hsize}{@{\extracolsep{\fill}}lccccc}
			\toprule[1pt]
			\midrule
			\multicolumn{1}{c}{User Posts} &
			Label &
			BERT &
			HAN &
			\begin{tabular}[c]{@{}c@{}}HAN-\\ MCM\end{tabular} &
			HSNPL(OURS) \\ \midrule
			\begin{tabular}[c]{@{}l@{}}1. I always keep blades in my wallet.\\ 2. wrist is burning, I love this feeling too much.\\ 3. this boy at my school cuts and gets suicidal.\\ 4. You're cheating on her. You don't love her.\\ 5. Cross the line if you feel totally worthless\\ and you have no idea why.\end{tabular} &
			positive &
			negative &
			negative &
			negative &
			positive \\ \midrule
			\begin{tabular}[c]{@{}l@{}}1. there are some people who only like the part of\\ your face that's smiling.\\ 2. I feel like an evil spirit just stopped staring at me.\\ 3. All wishes have a price.\\ 4. It hurts a lot but it's worth it.\\ 5. I cannot trust my phone anymore.\end{tabular} &
			positive &
			negative &
			negative &
			positive &
			positive \\ \midrule
			\begin{tabular}[c]{@{}l@{}}1. Bereaved families can plant snowdrops in\\ remembrance of loved ones at the annual walk.\\ 2. These essential oils can help you when dealing\\ with depression through a bereavement.\\ 3. 10 things I wish someone would have told me about \\ \#grief.\#bereavement \#counselling \#support.\\ 4. \#JohnTravolta posts touching tribute about his \\ \#bereavement after son's death.\\ 5. Coping with the loss of a loved one.\end{tabular} &
			positive &
			positive &
			positive &
			positive &
			positive \\ \midrule
			\begin{tabular}[c]{@{}l@{}}1. Today I feel that I want to learn and be fun and happy.\\ I LOVE TODAY.\\ 2. Tomorrow I hope it rains because I want to play \\ in the classroom with Rita.\\ 3. Today I do it science I'm doing it little be it help.\\ 4. Today I had P.E. boy vs. girl on doge ball castle \\ it was fun!\\ 5. Today I will have an airsoft gun Fusheng.\end{tabular} &
			negative &
			negative &
			negative &
			negative &
			negative \\  \midrule
			\bottomrule[1pt]
		\end{tabular*}
		\label{tab8}
	\end{table*}
	
	To demonstrate the advantages and effectiveness of our method, we select four representative cases from the dataset and show the prediction results of different methods in Table \ref{tab8}. The first example is $User4$ mentioned in Section \ref{Attention}, who is a depressed user but is misclassified as a non-depressed user by BERT, while in our method, he is correctly classified as a depressed user. To illustrate the effectiveness of our method, we sort and display the attention weights of the words in the user's tweets in descending order, as shown in Figure \ref{fig13}, where the word area is proportional to the attention weight. The attention weight reflects the word's contribution to feature representation for depression prediction. The results show that "chronic," "suicidal," and "clinical," which are related to depression, gain higher weights in our method, indicating that our method captures these words' influence on depression classification. At the same time, BERT is a generalized language model that cannot focus on task-oriented information effectively. However, when using HAN and HAN-MCM, this case is still misclassified. According to the analysis in Section \ref{Attention}, this user has a close social relationship with other users, and his depressive state can be revealed by other users' features and behaviors, which are not considered in BERT, HAN, and HAN-MCM. 
	
	In the second example, BERT and HAN mispredict the user because these models fail to capture the user's implicit psychological symbols in the tweets. At the same time, HAN-MCM utilizes linguistic knowledge to improve semantic understanding and correctly classifies this case. We map implicit psychological symbols via prompt learning, which can learn effective feature representations at the psychological level. Finally, the third and fourth cases contain prominent psychological symbols, expressing the user's depressed or non-depressed states, therefore, they are correctly classified by all comparison methods.
	
	\subsection{Effect of Anchor Tweet}
	\begin{figure*}[h]
		\centering
		\subfigure[Without anchored tweets]{
			\includegraphics[width=1.0\columnwidth]{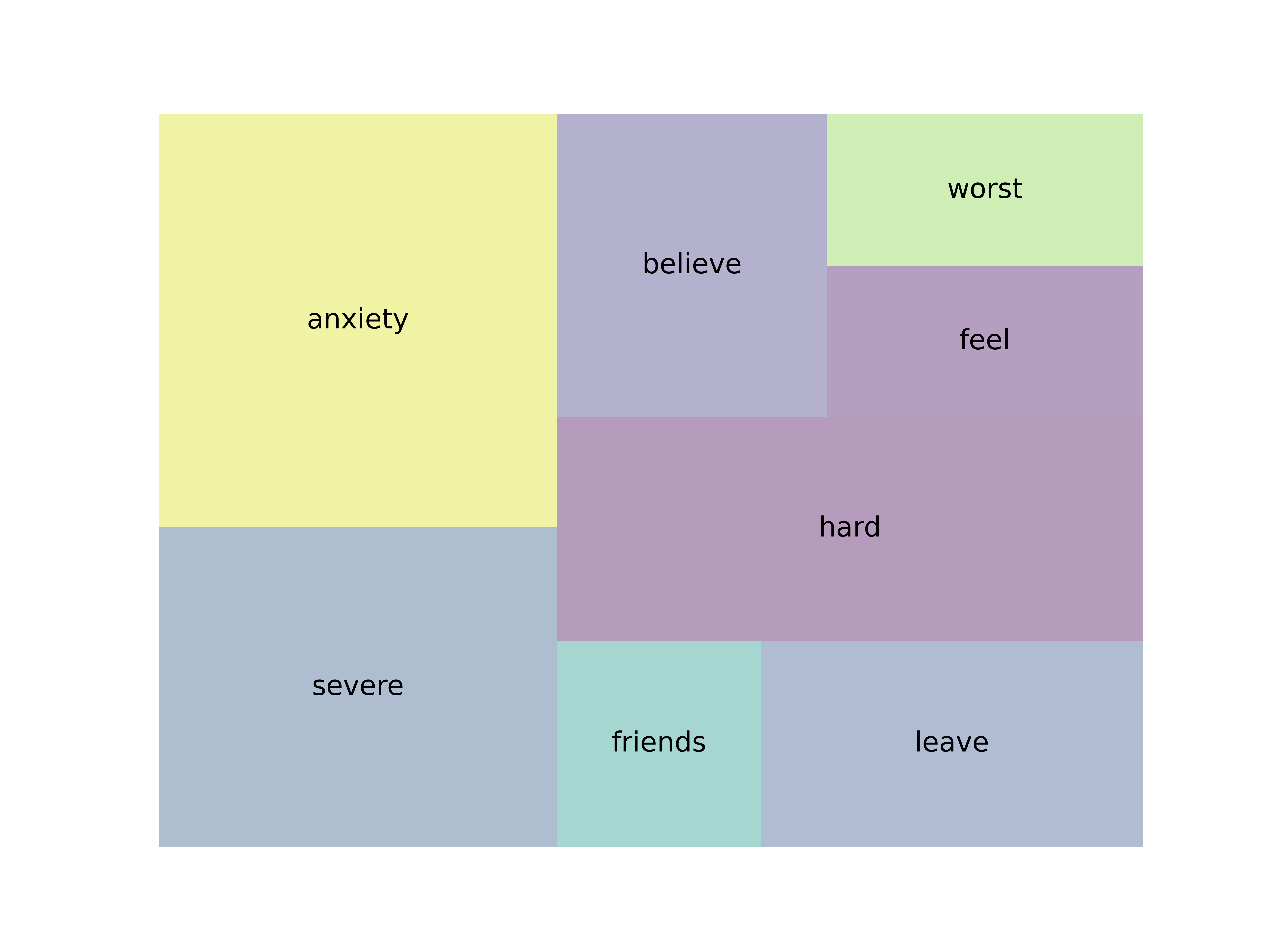}}
		\subfigure[With anchored tweets]{
			\includegraphics[width=1.0\columnwidth]{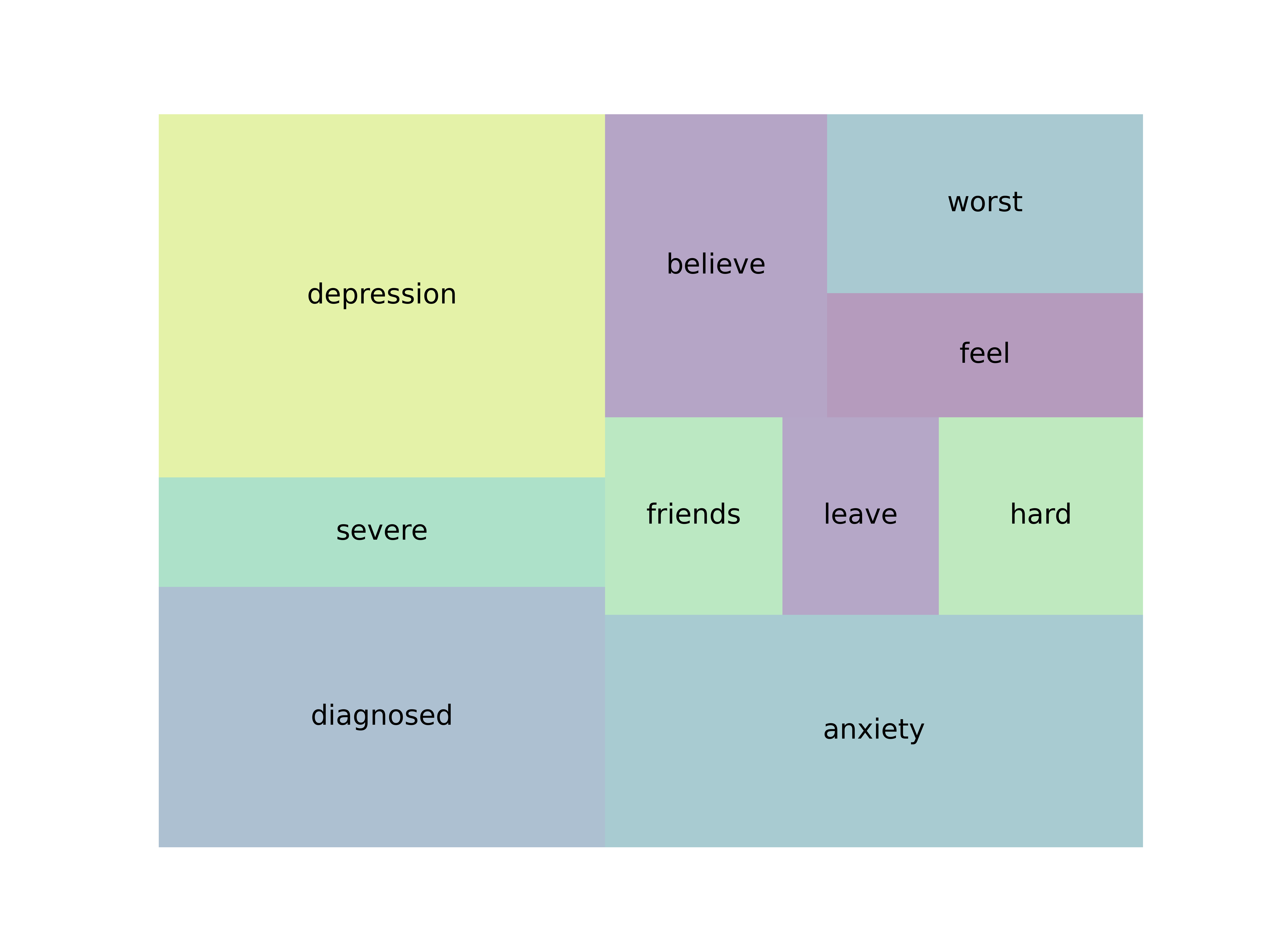}}
		\caption{Attention weights of words in anchor tweets.}
		\label{fig7}
	\end{figure*}
	
	We notice that Han et al. \cite{han2022hierarchical} performed their model HAN-MCM on the dataset without filtering out the anchor tweets. Therefore, we also conduct experiments on the same dataset with anchor tweets. As shown in Table \ref{tab1}, our method performs better under the condition with anchor tweets than HAN-MCM (improved by about 1$\%$), and the performance with anchor tweets is much better than that without anchor tweets, indicating that our method can effectively utilize the depressive symptoms in anchor tweets, which has a powerful guidance effect for depression detection. 
	
	To better comprehend the role of anchor tweets, we visualize the attention weights of words in anchor tweets of an example user under the conditions with/without anchor tweets in descending order, as shown in Figure \ref{fig7}. The results show that our method assigns higher attention weights to the keywords that are related to depression (such as "diagnosed" and "depression") in the anchor tweet under the condition with anchor tweets compared to without anchored tweets, indicating that the model pays more attention to them during feature learning. However, we usually do not have annotations similar to the anchor tweets in clinical diagnoses or applications. Therefore, excluding annotation information or anchor tweets in the dataset is more realistic and practical in the model training stage. Moreover, our method can still outperform the baselines, regardless of whether the anchor tweets are filtered out, demonstrating that our method is significantly effective for depression detection.
	
	\subsection{Error Analysis}
	\begin{table}[h]
		\centering
		\caption{Error analysis.}
		\resizebox{1.00\columnwidth}{!}{
			\begin{tabular}{lc}
				\midrule
				\multicolumn{1}{c}{User Posts} &
				Label \\ \midrule
				\begin{tabular}[c]{@{}l@{}}1. I only make people miserable.\\ 2. So sick of myself.\\ 3. Is it confusion, exhaustion or numbness? \\ I don't know too.\\ 4. I'm messed up.\\ 5. Worst is you're not around \&amp; \\ I can't talk to anyone.\end{tabular} &
				positive \\ \midrule
				\begin{tabular}[c]{@{}l@{}}1. Not even tryna get close to anybody.\\ 2. To the point now where If you wanna be\\ apart of my life then aye, it's lit.\\ 3. It's the small things that matter most.\\ 4. I hate the Internet.\\ 5. A Dream without execution is just a Dream.\end{tabular} &
				negative \\ \midrule
			\end{tabular}
		}
		\label{tab9}
	\end{table}
	
	Our investigation shows that some users (such as some overlapping points in Figure \ref{fig5.2}) posted abundant content related to specific topics, and the number of depressed and non-depressed users who posted these topics is imbalanced. For example, in a photography topic, only a few users were depressed users, but they posted content that was closely related to the photography topic. Therefore, the model sometimes cannot effectively distinguish them from other photography enthusiasts and misclassifies them as non-depressed users, indicating that our method needs to be improved in distinguishing users after learning the correlations between users. To solve this problem, in future work, we plan to develop adversarial learning to improve the model's sensitivity to different groups of users.
	
	On the other hand, we list some typical users who are incorrectly predicted in Table \ref{tab9}. Some depressed users (such as the first row in the table) are more inclined to personal behavior. Therefore, other users(especially non-depressed users) rarely mention their posted information and rarely share common topics with them. They also post significantly fewer tweets than the average number of tweets in the dataset, indicating their loneliness and social barriers associated with depression risk. Similarly, some non-depressed users who post less content are also incorrectly classified by the model (such as the second row in the table). This indicates that our method needs more generalization ability for users with sparse data. To solve this problem, in future work, we intend to leverage propagation chains to learn social relations and information propagation between users to enhance depression detection ability.
	
	\section{Conclusion}
	We propose a heterogeneous subgraph network incorporating prompt and contrastive learning strategies for interpretable depression detection. Implicit psychological symbols can be revealed by prompt learning at the psychological level with the aid of the depression scale, thereby the interpretability of the model is enhanced. Abundant information and complex relationships are captured by the heterogeneous graph structure constructed for multidimensional social data. Rich interactions among different types of information are explored at the feature level by a heterogeneous attention network with a dual attention mechanism. Latent interactions among social users and the group are explored by subgraph attention mechanism and self-supervised subgraph contrastive learning mechanism at the user level, resulting in more distinctive user representations. Our proposed model significantly improves the performance of depression detection on social media. 
	
	In future works, we will develop adversarial learning to enhance the sensitivity to different user groups and consider propagation chains to capture relationships between users to improve the generalization ability of the model for sparse data. Moreover, this depression detection model on social media primarily analyzes publicly available datasets and has not yet been deployed in practical applications. In future works, we will endeavor the cooperate with the psychological counseling center of our university for early depression detection of our students, which is urgent.
	
	\section{Ethical Considerations}
	This work is based on the dataset constructed and publicly released by Shen et al. \cite{shen2017depression}, which aims to predict the potential depressive states of users on social media. Therefore, this work does not require IRB/ethical approval. Moreover, we declare that we are against any misuse of our model in activities that violate data security, privacy protection, and ethics.
	
	\section{Acknowledgments}
	This work is supported by the Natural Science Foundation of Guangdong Province (No. 2021A1515012290), Guangdong Provincial Key Laboratory of Cyber-Physical Systems (No. 2020B1212060069), and National \& Local Joint Engineering Research Center of Intelligent Manufacturing Cyber-Physical Systems.
	
	
	

	
	
	

	
	\bibliographystyle{cas-model2-names}
	
	\bibliography{cas-refs}
	
	
	
\end{document}